\documentstyle[10pt,epsfig]{article}
\begin{document}

\title{Superfluid nuclear matter calculations}
\baselineskip=1. \baselineskip

\author{     P. Bo\.{z}ek$^\star$ \\
\centerline{ National Superconducting Cyclotron Laboratory,}\\
\centerline{ and Department 
of Physics and Astronomy,}\\ 
\centerline{Michigan State University, 
East Lansing, MI-48824} \\
\centerline{ and }\\
\centerline{Institute of Nuclear Physics, PL-31-342 Krak\'{o}w, Poland}}
\date{\today}
\maketitle 

\begin{abstract}
We present a method to calculate  nuclear matter properties in the
superfluid phase. The method is based on the use of  self-consistent off-shell
nucleon propagators in the T-matrix equation.
Such a complete treatment of the spectral function, is required below and 
around $T_c$ due to a pseudogap formation in the spectral function.
In the superfluid phase we introduce the anomalous self-energy in the fermion 
propagators and in the T-matrix equation, consistently with the 
strong coupling BCS equations.
The  equations for the nucleon
 spectral function include both a contribution of condensed  and 
scattering pairs.
The method is illustrated by numerical
 calculations.  Above $T_c$ pseudogap formation 
is visible in the spectral function   and below $T_c$ a superfluid gap also
appears.  
\end{abstract}



\vspace{1cm}
\noindent
$^\star${electronic
address~:~bozek@solaris.ifj.edu.pl}

\section{Introduction}

Nuclear matter calculations in the Brueckner approximation \cite{nm,fw}
do not consider 
a possible pairing in the ground state  of nuclear matter \cite{sfnm},
with the exception of Refs. \cite{d2}. 
This approach is justified by the fact that the superfluid energy gap is 
expected to be small in normal density nuclear matter \cite{fw}.
When addressing the spectral properties of nucleons in medium, calculations
using T-matrix approximation where performed \cite{vo,d1,ko,roepke}.
The T-matrix approximation for the two-particle Green's functions leads 
to a much stronger pairing than the Brueckner approximation. The T-matrix 
calculation and the BCS theory give the same prediction for the critical 
temperature \cite{thouless}. Brueckner type calculations of the 
effective in-medium nuclear interactions, permit to estimate  approximate 
nuclear  matter ground state energy , in-medium nucleon masses,
 and effective cross 
sections from the free nucleon-nucleon interaction. Although, discrepancies 
between different calculations and different parameterization of the 
nuclear interaction remain, a
 tremendous improvement over  a simple Hartree-Fock approach
is achieved. It is an interesting question, how to include the pairing force 
into this picture. 

The generalization of the T-matrix nuclear matter calculations below the 
superfluid transition temperature ($T_c$), requires by itself a procedure
combining the ladder summation of the self-energy diagrams and a possible
fermion pairing. On the other hand, the T-matrix scheme seems to be a natural
starting point for the inclusion of superfluidity in nuclear matter 
calculations \cite{thouless,km}. In this work we show that such a procedure
is indeed possible, merging a T-matrix calculation of the normal fermion
self-energy with a mean-field, frequency independent, superfluid energy gap.
The interaction potential, which enters in the superfluid gap equation 
is the free nucleon-nucleon interaction. However, the scattering of nucleons 
and the pseudogap formation above the critical temperature reduces
the value of the energy gap and of the  critical temperature, as 
compared to the BCS result.
This can be achieved only using self-consistent off-shell propagators
in the T-matrix diagrams \cite{ja}. In such a way the formation of a pseudogap 
influences strongly the result for the T-matrix in attractive channels, 
shifting the appearance of a singularity in the T-matrix
 at the Fermi energy to much lower 
temperatures. A consistent treatment of the superfluid transition requires 
then  both the use of the off-shell nucleon propagators  
(below and above $T_c$) and of the anomalous self-energy (below $T_c$).
To illustrate the formation of the pseudogap above $T_c$, we present first the
 results of two calculations  above $T_c$ (Sec. \ref{pseudosec}),
 a  self-consistent T-matrix resummation, using off-shell propagators and a
T-matrix calculation using on-shell nucleon propagators with Hartree-Fock 
single-particle energies. The results of that section indicate the necessity
of using  off-shell propagators in the T-matrix equation
around  $T_c$, rather 
than a quasi-particle approximation.
In section \ref{pairing} we derive the equation for the T-matrix
self-energy  and the gap equations, in the superfluid phase.
A numerical example of a self-consistent calculation of the 
superfluid energy gap
and of the T-matrix self-energy is presented is section \ref{super}.

\section{T-matrix equation in the Fermi liquid phase}

The nuclear matter can be described as an infinite system of interacting
 fermions. We consider a system of neutrons and protons interacting via
two-body forces
\begin{eqnarray}
H=\sum_\alpha \int d^3 x \ \Psi_\alpha^\dagger(x)(-\frac{\Delta^2}{2 m}-\mu)
\Psi_\alpha(x) \nonumber \\
 + \sum_{\alpha^{'},\beta^{'},\alpha,\beta}
\frac{1}{2}\int d^3 x \int d^3 y \Psi_{\alpha^{'}}^\dagger(x) 
\Psi_{\beta^{'}}^\dagger(y)  V_{\alpha^{'},\beta^{'},\alpha,\beta}(x,y)
\Psi_{\beta}(y) 
\Psi_\alpha(x) \ .
\end{eqnarray}
Note that we define the energies with respect to the Fermi energy 
(chemical potential) $\mu$; 
also for
  real-time calculations. This will be helpful when dealing with the 
anomalous propagator in the superfluid phase.
In the following we shall use equivalently the term zero energy and the Fermi 
energy.
Nuclear forces can decomposed in  (JST) (total angular-momentum, spin, isospin)
channels \cite{gw}
\begin{equation}
V_{\alpha^{'},\beta^{'},\alpha,\beta}({\bf k},{\bf p}) = \sum_{(JST)ll^{'}}
 i^{l-l^{'}} 
{\cal Y}^{(JST)l}_{\alpha^{'}\beta^{'}}({\hat k})
 V^{(JST)}_{ll^{'}}(k,p) {\cal Y}^{(JST)l^{'}}_{\alpha \beta}({\hat p})^\star
 \ .
\end{equation}
The sum over the partial waves $l,l^{'}$ is restricted to $l=l^{'}=J$ for
uncoupled states and to  $l,l^{'}=J\pm 1$ for the scattering of
 triplet states.

One body observables of a systems of fermions at equilibrium can be 
described by the spectral function
\begin{eqnarray}
\label{spectral}
A(p,\omega)&=&-2{\rm Im} G^+(p,\omega) \nonumber \\
&=&\frac{-2{\rm Im}\Sigma^{+}(p,\omega)}{
(\omega-p^2/2 m-{\rm Re} \Sigma^+(p,\omega)+\mu)^2+{\rm Im}
\Sigma^{+}(p,\omega)^2 } \ ,
\end{eqnarray}
where $G^\pm$ denote the retarded(advanced) Green's function.
We restrict the discussion to symmetric nuclear matter so that the Green's
function is diagonal in  spin-isospin indices.
The T-matrix approximation consist of summing  particle-particle 
and hole-hole ladder diagrams for the self-energy \cite{kb,paw1,bm}.
The self-energy for in-medium propagators describes the dressing of the 
quasi-particles due to scatterings with other particles. In general one cannot 
approximate the spectral function by a $\delta$-function (quasi-particle peak) 
at some quasi-particle 
energy $\zeta_k$. The particles acquire a width, which enters 
self-consistently in the equation for the T-matrix.
Moreover,  more complicated 
structures in the spectral function seem to appear when the temperature 
is in a range of few MeV above the superfluid transition \cite{roepke}.
Using a quasiparticle ansatz for the spectral-function the authors of Ref.
\cite{roepke} have shown that the spectral function 
has two  (or sometimes three)
peaks. This effect by itself invalidates the 
quasi-particle approximation for the T-matrix.
On the other hand, it is possible to
 calculate the T-matrix and the corresponding spectral-function 
self-consistently using off-shell propagators in the ladder diagrams
for the self-energy \cite{ja}. The self-consistent calculation was performed in
 a simple model with separable interactions in the $S$-wave. In the present
work we use  this simple interaction for numerical examples.
The self-consistent solution encompasses in a consistent way the non-Fermi 
liquid behavior of the spectral functions, and, thus, gives quantitatively 
different results for the critical temperature and for the temperature where
 the pseudogap appears, than the quasi-particle approximation.

Off-shell propagators were used previously in a nuclear matter 
calculation by Jong and Lenske \cite{JL}.
 The spectral function in that work was approximated as a sum of a 
quasiparticle peak and a continuum part on the other side of the Fermi energy,
the contribution to the two particle propagator with two continuum spectral
functions was neglected to simplify the numerics. This neglected part could be
calculated using the methods developed in the present work. 
On the other hand a mixed ansatz using a quasi-particle and a continuum part
for the spectral function, like the one used in \cite{JL} or similar,
 is unavoidable at small temperatures.
The authors of Ref \cite{JL} use a self-consistent scheme a zero temperature,
 but
do not consider pairing. However, even if the expected pairing is small
it is more natural to use a scheme allowing for the creation of pair 
condensate, since the normal Fermi  liquid  
is unstable.

The T-matrix   for a system with a
two-body interaction $V({\bf p},{\bf p}^{'})$ is defined as 
\cite{kb,paw1,bm,schmidt}~:
\begin{eqnarray}
\label{teq}
<{\bf p}|T_{\alpha^{'}\beta^{'}\alpha\beta}
^\pm({\bf P},\omega)|{\bf p}^{'}>& =& V_{\alpha^{'}\beta^{'}\alpha\beta}
({\bf p},{\bf p}^{'}) \nonumber \\ & & + \sum_{\gamma \delta}
 \int\frac{d^3k}{(2 \pi)^3}
\int\frac{d^3q}{(2 \pi)^3} V_{\alpha^{'}\beta^{'}\gamma\delta}
({\bf p},{\bf k}) \nonumber \\ 
& &<{\bf k}|{\cal G}^\pm({\bf P},\omega)|{\bf q}>
 <{\bf q}|T_{\gamma\delta\alpha\beta}^\pm({\bf P},\omega)
|{\bf p}^{'}> \ ,
\end{eqnarray}
where the disconnected two-particle propagator is~:
\begin{eqnarray}
\label{twpro}
<{\bf p}|{\cal G}^\pm({\bf P},\omega)|{\bf p}^{'}> =  
(2 \pi)^3 \delta^3({\bf p}-{\bf p}^{'})\int \frac{d\omega^{'}}{2 \pi}
\int \frac{d\omega^{''}}{2 \pi} \nonumber \\
\Big( G^<({\bf P}/2+{\bf p},\omega^{''}-\omega^{'})G^<({\bf P}/2-{\bf p},
\omega^{'}) \nonumber \\ -G^>({\bf P}/2+{\bf p},\omega^{''}-
\omega^{'})G^>({\bf P}/2-{\bf p},\omega^{'}) \Big)/ 
\Big(\omega -\omega^{''} \pm 
i\epsilon\Big)  \ ,
\end{eqnarray}
with
\begin{eqnarray}
G^{<}(p,\omega)& =&  i A(p,\omega) f(\omega) \nonumber \\
G^{>}(p,\omega)& =&  -i A(p,\omega) \Big(1-f(\omega)\Big)  \ ,
\end{eqnarray}
and
\begin{equation}
f(\omega)=\frac{1}{e^{\omega/T}+1} 
\end{equation}
is the Fermi distribution.
The in-medium T-matrix equation can be decomposed in partial waves 
if an angle averaged two-particle propagator $\langle {\cal G} \rangle_\Omega$
is used~:
\begin{equation} 
\langle <{\bf p}|{\cal G}^\pm({\bf P},\omega)|{\bf p}^{'}> \rangle_\Omega
=\int \frac{d \Omega}{4 \pi} <{\bf p}|{\cal G}^\pm({\bf P},\omega)|{\bf p}^{'}>
\ ,
\end{equation}
where we average over the angle between ${\bf P}$ and ${\bf p}$.
\begin{eqnarray}
\label{teqp}
<{p}|T^{(JST) \ \pm}_{l^{'}l}
({ P},\omega)|{ p}^{'}>& =& V^{(JST)}_{l^{'}l}
({ p},{ p}^{'})+ \sum_{l^{''}}
 \int\frac{k^2 d k}{(2 \pi)^3}
 V_{l^{'}l^{''}}^{(JST)}
({ p},{ k}) \nonumber \\ 
& &<{ k}|{\cal G}^\pm({ P},\omega)|{ k}>
 <{ k}|T^{(JST) \ \pm}_{l^{''}l}({ P},\omega)
|{ p}^{'}> \ ,
\end{eqnarray}

In the course of this work we shall discuss two different
approximation schemes for the intermediate 
two-particle propagator in the T-matrix calculation.
The formulas for the in medium T-matrix approximation in nuclear matter with 
partial wave decomposition can be found in Ref. \cite{schmidt}.

\begin{itemize}
\item{\bf Self-consistent solution}

 The self-consistent solution uses  off-shell nucleon 
propagators in the ladder diagrams (Eq. \ref{teq}). The imaginary part of the
self-energy is defined by the 
T-matrix in the following way 
\begin{eqnarray}
\label{imsc}
{\rm Im}
\Sigma^+(p,\omega) = \frac{1}{8 \pi}\sum_{(JST)l} (2T+1)(2J+1)
\int\frac{d\omega^{'}}{2 \pi}\int \frac{d^3k}{(2 \pi)^3}
 \nonumber \\  <({\bf p}-{\bf  k})/2|{\rm Im}T^{(JST) \ +}_{ll}
(|{\bf p}+{\bf k}|,\omega+\omega^{'})|({\bf  p}-{\bf k})/2>
\nonumber \\
 A(k,\omega)\Big( f(\omega^{'})+b(\omega+\omega^{'}) \Big) \ ,
\end{eqnarray}
 where
\begin{equation}
b(\omega)=\frac{1}{e^{\omega/T}-1} 
\end{equation}
is the Bose distribution.
The real part of the self-energy is the sum 
\begin{equation}
\label{realpart}
{\rm Re} \Sigma(p,\omega)=\Sigma_{HF}(p)+ \Sigma_{d}(p,\omega)
\end{equation}
of the Hartree-Fock term
\begin{eqnarray}
\label{hfsc}
\Sigma_{HF}(p)=\frac{1}{8 \pi}\sum_{(JST)l} (2T+1)(2J+1)\int
\frac{d\omega}{2 \pi}\int \frac{d^3k}{(2 \pi)^3} \nonumber \\
V^{(JST)}_{ll}\big(
|{\bf p}-{\bf k}|/2,|{\bf p}-{\bf k}|/2\big) 
A(k,\omega) f(\omega) \ ,
\end{eqnarray}
and the dispersive contribution to real part of the self-energy
\begin{equation}
\label{disc}
\Sigma_{d}(p,\omega)={\cal P}
\int \frac{d \omega^{'}}{ \pi} \frac{- {\rm Im} \Sigma^+
(p,\omega^{'})}{\omega-\omega^{'}} \ .
\end{equation}
The  solution of Eqs. (\ref{teqp}), (\ref{imsc}), (\ref{hfsc}), (\ref{disc}),
and (\ref{spectral})
is obtained by iteration with the constraint
\begin{equation}
\label{const}
4 \int \frac{d \omega}{2 \pi}\int \frac{d^3 p}{(2 \pi)^3}A(p,\omega)
f(\omega)=\rho \ ,
\end{equation}
where $\rho$ is the assumed density of the nuclear matter.
Numerical methods used in the solution of the self-consistent T-matrix
 equations are presented in an appendix.

\item{\bf Quasi-particle approximation} 

The quasi-particle solution is obtained 
by using  mean-field propagators in the T-matrix equation.
\begin{equation}
\label{mfpro}
G_{mf}^\pm(p,\omega)=\frac{1}{\omega-\xi_p\pm i \epsilon} \ ,
\end{equation}
where $\xi_p=\omega_p-\mu$ is the mean-field energy measured with respect 
to the Fermi energy, and the mean-field quasi-particle energy is of course
$\omega_p=p^2/(2 m)-\Sigma_{HF}(p)$.
More generally, the self-consistent real part of the self-energy 
(\ref{realpart}) can also be used to define the quasi-particle energy.
The density is given by 
\begin{equation}
\rho=4 \int \frac{d^3 p}{(2 \pi)^3} 
f(\xi_p) \ .
\end{equation}
Analogously the expressions for the self-energies in the quasi-particle
approximation take the form
\begin{eqnarray}
\label{simfqp}
\Sigma_{HF}(p)=\frac{1}{8 \pi}\sum_{(JST)l} (2T+1)(2J+1)
\int \frac{ d^3k}{(2 \pi)^3} \nonumber \\ 
V^{(JST)}_{ll}\big(
(|{\bf p}-{\bf k}|/2,|{\bf p}-{\bf k}|/2\big)
 f(\xi_k) \ ,
\end{eqnarray}
\begin{eqnarray}
\label{imqp}
{\rm Im}
\Sigma^+(p,\omega) = \frac{1}{8 \pi}\sum_{(JST)l} (2T+1)(2J+1)
\int \frac{d^3k}{(2 \pi)^3}
 \nonumber \\  <({\bf p}-{\bf k})/2|{\rm Im}T^{ (JST) \ +}_{ll}(|{\bf p}
+{\bf k}|,\omega+\xi_k)|({\bf  p}-{\bf  k})/2>
\nonumber \\
 \Big( f(\xi_k)+b(\xi_k+\omega) \Big) \ .
\end{eqnarray}
The dispersive contribution to the real part of the self-energy can be
 obtained  from the dispersion relation (\ref{disc}).
Note that there is no need for an iteration procedure in the solution of 
the T-matrix equation in the quasi-particle approximation, when using 
Hartree-Fock single particle energies, as in this work. 
Once  the mean-field self-energy is calculated,
 the T-matrix can be obtained as
\begin{eqnarray}
\label{teqqp}
<{ p}|T^{ (JST) \ \pm }_{l^{'}l}({ P},\omega)|{ p}^{'}>=
 V^{(JST)}_{l^{'}l}({ p},{ p}^{'}) +\sum_{l^{''}}  
\int\frac{k^2dk}{(2 \pi)^3} V^{(JST)}_{l^{'}l^{''}}({ p},{ k})
\nonumber \\
 \frac{\langle 1-f(\xi_{p_1})-f(\xi_{p_2})\rangle_{\Omega}}{\omega
-\langle (\xi_{p_1}+\xi_{p_2})\rangle_{\Omega}
\pm i \epsilon} 
 <{k}|T^{(JST) \ \pm}_{l^{''}l}({ P},\omega)
|{ p}^{'}> \ , \\
{\bf p_{1,2}}={\bf P}/2\pm{\bf k} \ . \nonumber
\end{eqnarray}
For numerical convenience,
 we calculate the angle average separately in the numerator and 
in the denominator of the two-particle propagator in the above equation.
The spectral function is then calculated using Eq. (\ref{imqp}),
 the dispersion relation (\ref{disc}), and the definition (\ref{spectral}).
If the spectral function presents a quasi-particle peak around the energy 
$\xi_p$, the quasi-particle approximation can be trusted.

Of course an iterative improvement of the single particle energies 
using the real part of the self-energy obtained in the T-matrix or G-matrix 
approximation is possible \cite{d2,vo}.
Such an iterative scheme requires in principle  the introduction
of BCS-like single particle energies in the presence of pairing. 
This can be calculated in a quasi-particle version of the
formalism used in Sect. \ref{pairing}. However the actual implementation of
this approach is quite tedious and we restrict ourselves to the Hartree-Fock
single particle energies and to the normal Fermi-liquid phase for the the
quasi-particle approximation.

\end{itemize}

A common property of all the T-matrix approaches to the nuclear matter is the
appearance of strong pairing in channels with attractive interactions.
 As the temperature
 is lowered a singularity builds up in the T-matrix at zero total momentum 
and energy. For the quasi-particle scheme it leads to the Thouless criterion 
for $T_c$, which is equivalent to weak coupling BCS \cite{thouless}.
 The inclusion of
 self-energy corrections in the normal fermion propagators generalizes this
criterion, giving more general  equations for the critical temperature.
In  section \ref{pairing} we shall derive equations valid also below $T_c$.

\section{Pseudogap formation in finite temperature nuclear matter}
\label{pseudosec}

Numerical results presented in this, and subsequent sections were obtained 
using a Yamaguchi separable potential of rank one \cite{yama}.
 The momenta of all
 the nucleons are restricted to $|p|<700$~MeV. The coupling strengths 
of the original Yamaguchi parameterization are 
slightly adjusted to reproduce the scattering length in vacuum as obtain 
from the Yamaguchi potential without cutoff
in momenta of nucleons
\begin{eqnarray}
V(k,p)=\sum_{\alpha=^1S_0,^3S_1} \lambda_\alpha g(k)g(p) \ , \nonumber \\
g(p)=\frac{1}{p^2+\beta^2} \ ,
\end{eqnarray}
with
$\lambda_{^1S_0}=- .841 \  {\rm GeV}^2 $,
$\lambda_{^3S_1}=- .8591 \  {\rm GeV}^2$, and $\beta =285.9 \ {\rm MeV}$~.
For large momenta the effect of the cutoff is significant. Nevertheless,
the following calculation is a helpful illustration, presenting all the
basic phenomena related to the superfluidity in nuclear matter, and 
describing qualitatively 
the behavior around the Fermi energy. 
The binding energy of the deuteron and the critical temperature 
(in the quasi-particle approximation) are changed 
within few percent by the introduction of the cutoff.

All the results are calculated for
nuclear density $\rho=.45 \rho_0$. 
At higher densities, a resonance appears in the two-particle Green's function
below the Fermi energy. This happens for temperatures above $T_c$,
 according to the  Thouless criterion.
 This statement is strongly dependent on the assumed interaction
and the resulting deuteron wave function.
 This phenomenon is not related to
the main subject of this work.
 We postpone the discussion of this topic,
using more realistic parameterizations of the momentum distribution
of the deuteron, to
another work and restrict ourselves to low density nuclear matter in the
 following. The Yamaguchi potential has also the practical advantage, that
the critical temperature and the energy gap are relatively 
(somewhat unrealistically)
 high \cite{roepke}.
The strength of the pairing  simplifies the numerics, and is helpful
in  illustrating qualitatively the physical effects.
In Ref. \cite{ja} we have studied the self-consistent T-matrix 
approximation at normal nuclear density and high temperature. 
However we were not able to perform similar calculations at small temperatures
(around and below the critical temperature)
with the limited resolution in energy, because the single-particle
width becomes very small at the Fermi energy (See appendix).

\subsection{Quasi-particle approximation}

\begin{figure}
\begin{minipage}[t]{0.48\linewidth}
\centering
\epsfig{file=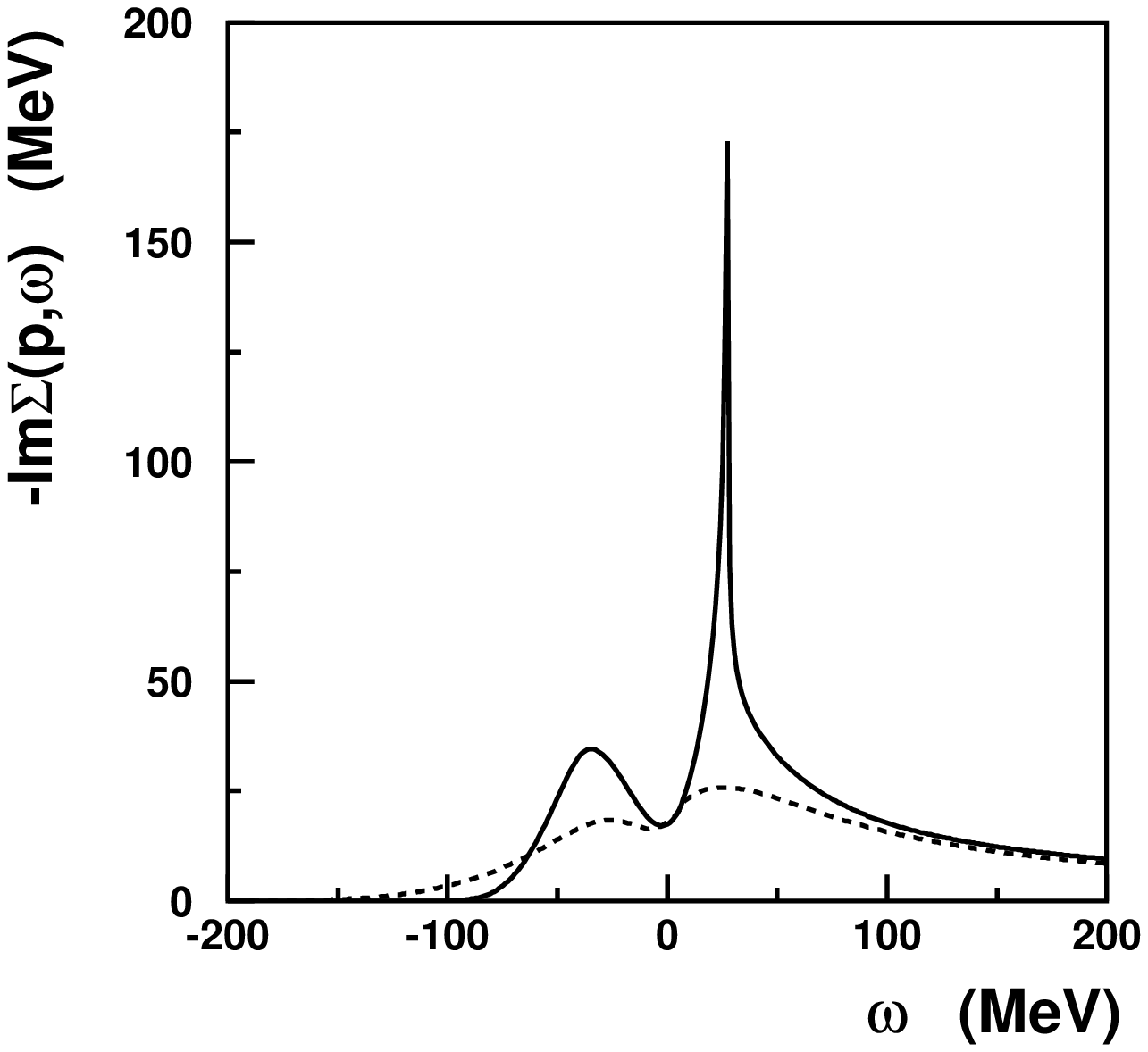,width=0.95\textwidth}
\caption{Imaginary part of the self-energy $-{\rm Im}\Sigma(p,\omega)$
as function of the energy for $p=0$ and $p=175$~MeV 
(solid and dashed lines respectively), calculated in the quasi-particle 
approximation for the T-matrix at $\rho=.54\rho_0$, and $T=5.1$~MeV
(slightly above $T_c$=5.03).}
\label{gamqp}
\end{minipage}
\hspace{.1in}
\begin{minipage}[t]{0.48\linewidth}
\centering
\epsfig{file=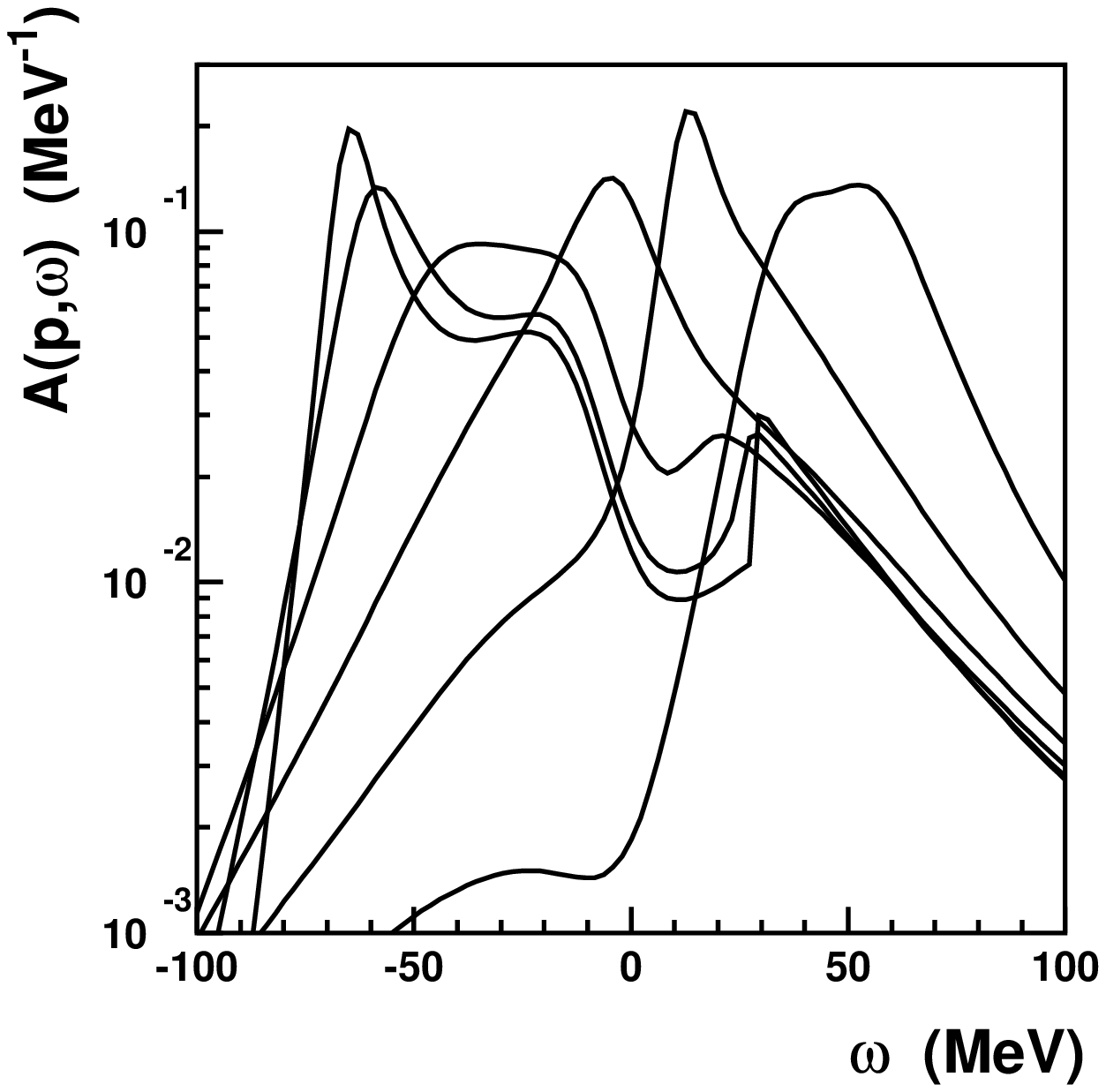,width=0.95\textwidth}
\caption{Spectral function  $A(p,\omega)$ for several values of momentum as 
function of energy for
the same parameters as in Fig. \ref{gamqp}.
 The lines correspond to $p=0, 70, 140, 210$ and $350$ MeV,
 according to the position, from left to right, of the largest peak
 in the spectral function.}
\label{spec51qp}
\end{minipage}
\end{figure}

The spectral function in the quasi-particle approximation has been calculated 
in several works \cite{vo,roepke,baldo,www}, in the T-matrix or Brueckner 
schemes. 
Calculations using the T-matrix 
approximation for the fermion self-energy show very strong pairing in 
channels with attractive interactions.
At a critical temperature a pole appears in the T-matrix. For the interaction
here chosen, it appears first in the  in the $^{3}S_{1}$ channel.
It signals  the formation of neutron-proton bound states 
(with zero binding energy, since the pole in the T-matrix appears 
at zero total momentum and at the Fermi energy) \cite{thouless,roepke},
 for even lower temperature it is preferable for the system to condense into 
nucleon-nucleon pairs instead of remaining in the Fermi-liquid phase.
At temperatures slightly above $T_c$, the T-matrix is strongly
peaked for small values of total momentum and small energy. It is a 
precursor of the  singularity in the T-matrix at the 
critical temperature. One of the consequences of this singularity is a very
 unusual form of the imaginary part of the self-energy 
(Fig. \ref{gamqp}). The imaginary part of the self-energy ${\rm Im}
\Sigma^{+}(p=0,\omega)$ has a peak for $\omega=-\xi_{p=0}$. Not surprisingly,
 the spectral function exhibits also  a nontrivial structure, with two
 or more peaks (Fig. \ref{spec51qp}). The spectral function is not of a 
Fermi-liquid type.

\begin{figure}
\begin{minipage}[t]{0.48\linewidth}
\centering
\epsfig{file=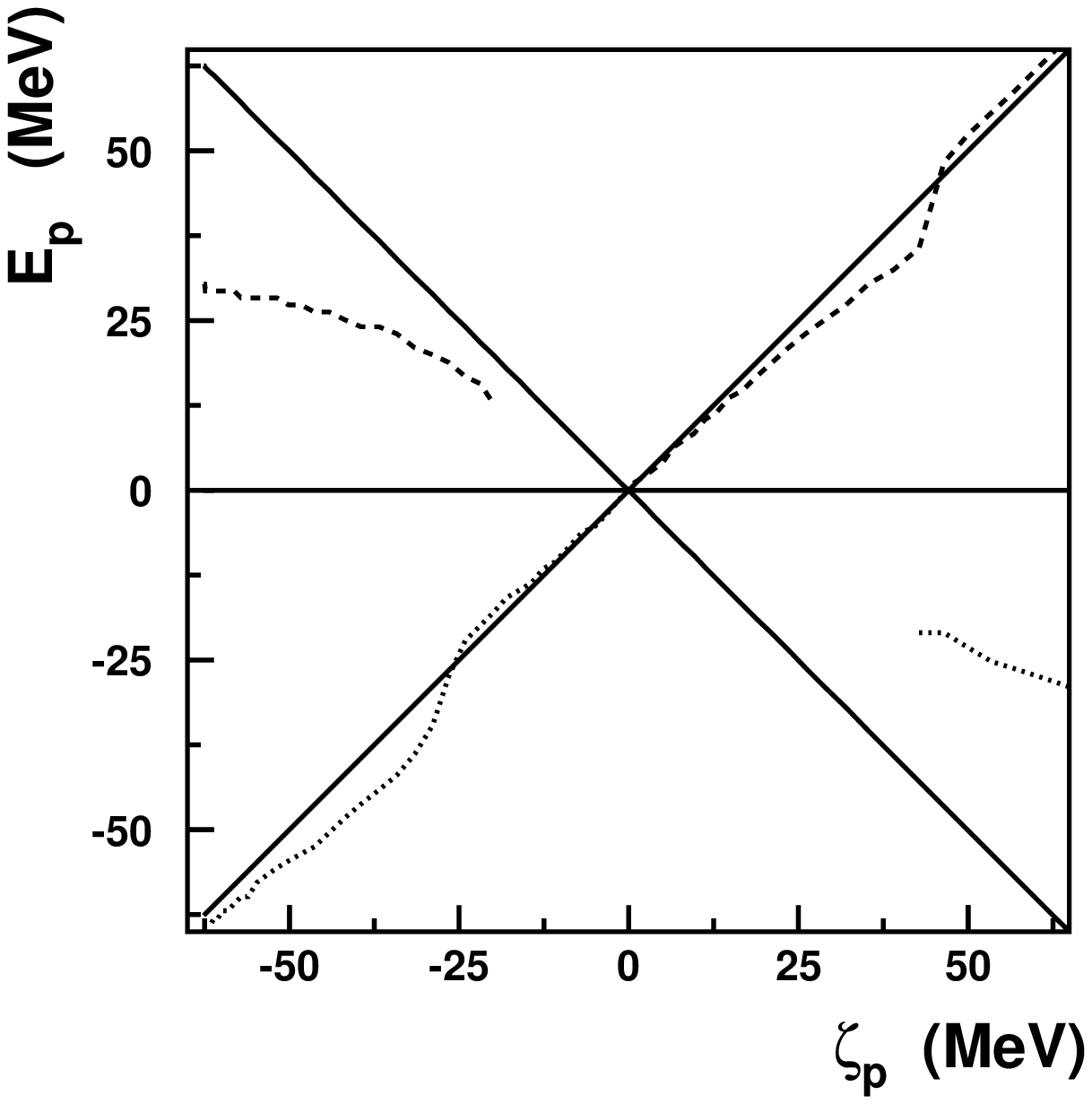,width=0.95\textwidth}
\caption{The positions of the largest  peaks of the spectral function
at positive and negative energies
(dotted and dashed lines) as function of the energy of the 
 quasi-particle pole,
 for the same parameters as in Fig. \ref{gamqp}.
 The solid lines denote the two asymptotic branches in the 
BCS solution $E_p=\pm \zeta_p$ and the Fermi energy $E_p=0$.}
\label{gap51qp}
\end{minipage}
\hspace{.1in}
\begin{minipage}[t]{0.48\linewidth}
\centering
\epsfig{file=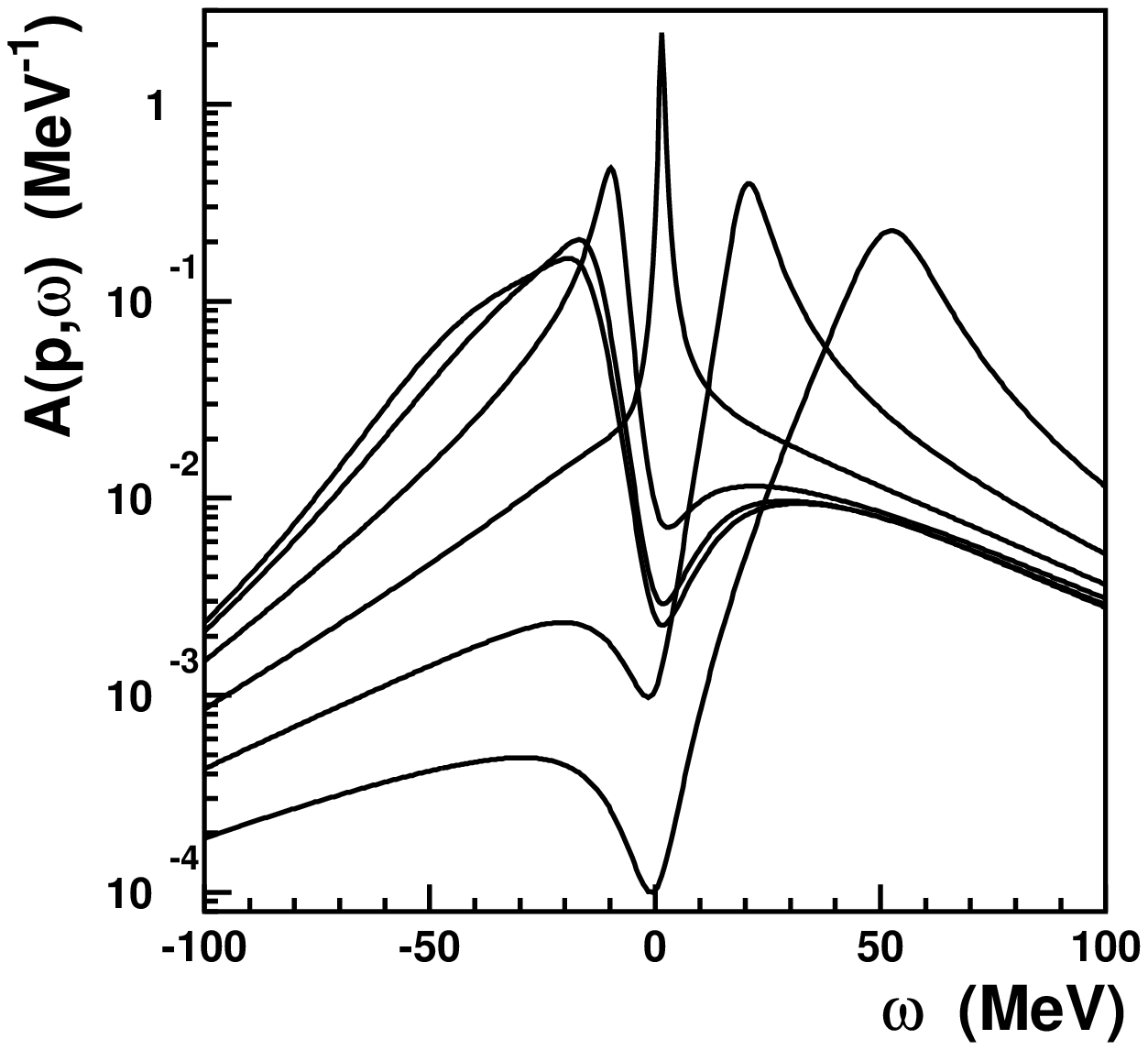,width=0.95\textwidth}
\caption{Spectral function for several values of momentum as 
function of energy.
 The lines correspond to $p=0, 70, 140, 210$ and $350$ MeV,
 according to the position, from left to right, of the largest peak
 in the spectral function.
The results were obtained 
in the self-consistent scheme at $T=1.7$ MeV  (slightly above $T_c=1.63$~MeV)
and $\rho=.45\rho_0$.}
\label{spec17}
\end{minipage}
\end{figure}

The appearance of two peaks in the spectral function
above $T_c$ was broadly discussed in high $T_c$ superconductivity 
under the name of pseudogap formation
\cite{pseudocm,peder,chica1,chica3,sccm,hau2}.
The quasi-particle approximation represents only the first iteration in the
 self-consistent calculation of the nucleon spectral function. Thus, it fails
to describe the feedback of the pseudogap on the value of the T-matrix in the
 vicinity of the Fermi energy.
 This  leads to quantitative difference between the self-consistent 
and the quasi-particle calculations.
The critical temperature is $1.63$ MeV for the self-consistent calculation and
$5.03$ MeV in the quasi-particle approximation. 
The form of the spectral function is different in the 
two approximations \cite{ja}.
The strong differences between the quasi-particle approximation and the 
self-consistent solution indicate that for strongly interacting systems
(having bound or almost bound states in vacuum) the usual approximation
 of the spectral function by a quasi-particle peak fails already above $T_c$.
Another difference already mentioned in \cite{ja} is that the imaginary 
part of the self-energy has a singularity in the quasi-particle approximation;
 in the self-consistent calculation this singularity is smeared 
out. In Fig. \ref{gap51qp} we plot the position of the maxima of the 
peaks in the spectral function 
as  function of the position of the quasi-particle pole
 $\zeta_p$~; 
\begin{equation}
\label{sieq}
\zeta_p-\frac{p^2}{2m}-{\rm Re} \Sigma(p,\zeta_p)+\mu=0 \ .
\end{equation}
Clearly two peaks are visible in the spectral function on
 two sides of the Fermi energy. The dominant peak follows approximately the 
position of the quasi-particle pole. However, a subdominant peak is 
found on the other side of the Fermi energy, recalling the effect of an
 energy gap in the superfluid phase. 
The effect was named a pseudogap, since the energy gap in the spectral 
function occurs without pairing condensate and the superfluid density is 
zero. 
Close to the Fermi energy the subdominant peak disappears.
The position of the quasi-particle peak (solution of Eq. \ref{sieq})
is relatively smooth as a function of momentum, in both the quasi-particle
 approximation and in the self-consistent solution. There is no sharp
variation of the effective mass near the Fermi energy.

\subsection{Self-consistent solution}

\label{pseudosecsc}

\begin{figure}
\begin{minipage}[t]{0.48\linewidth}
\centering
\epsfig{file=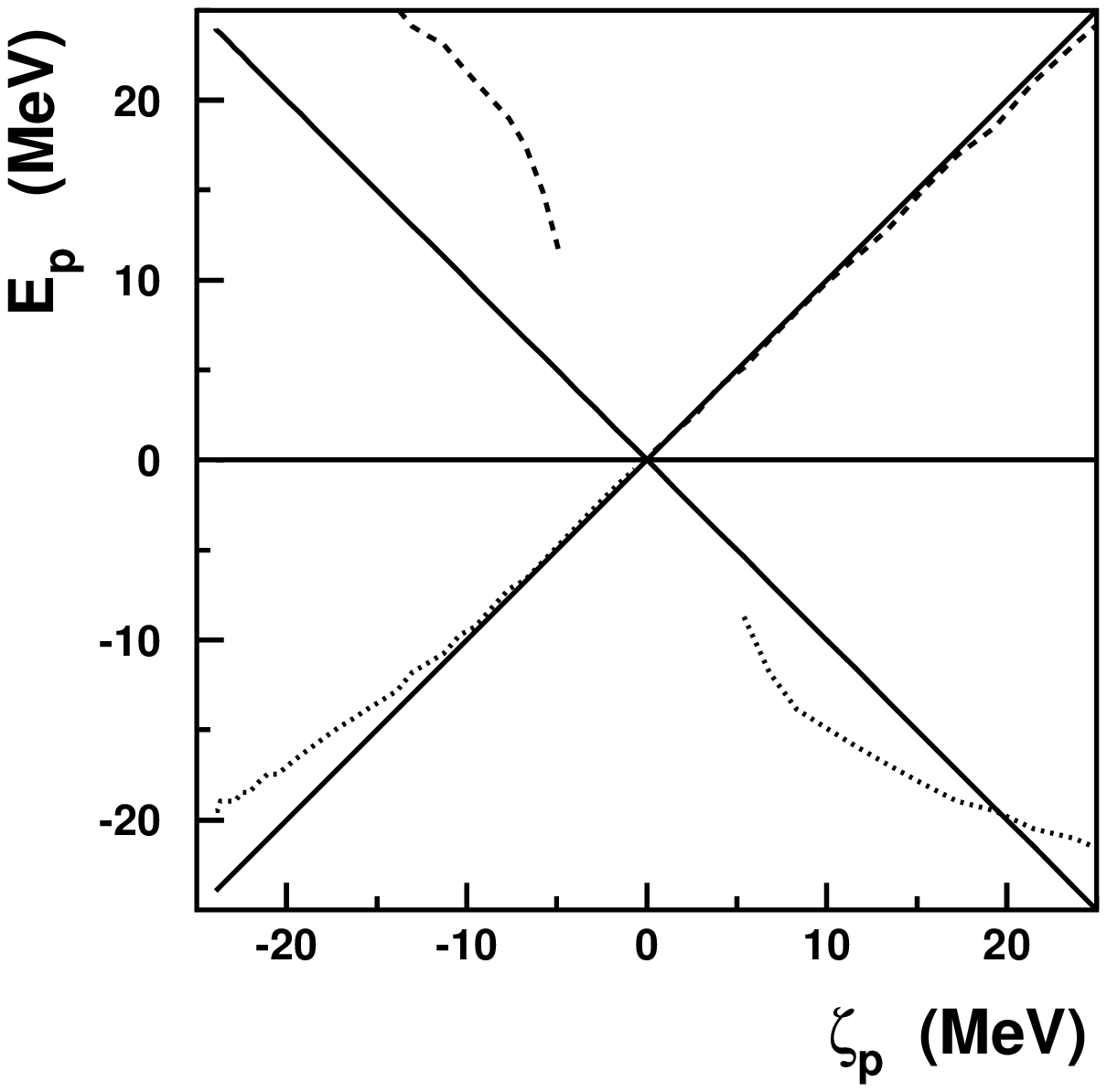,width=0.95\textwidth}
\caption{The positions of the two peaks of the spectral function
(dotted and dashed lines) as function of the energy of the 
quasi-particle pole,
for the same parameters as in Fig. \ref{spec17}.
 The solid lines denote the two asymptotic branches in the 
BCS solution $E_p=\pm \zeta_p$ and the Fermi energy $E_p=0$.}
\label{gap17}
\end{minipage}
\hspace{.1in}
\begin{minipage}[t]{0.48\linewidth}
\centering
\epsfig{file=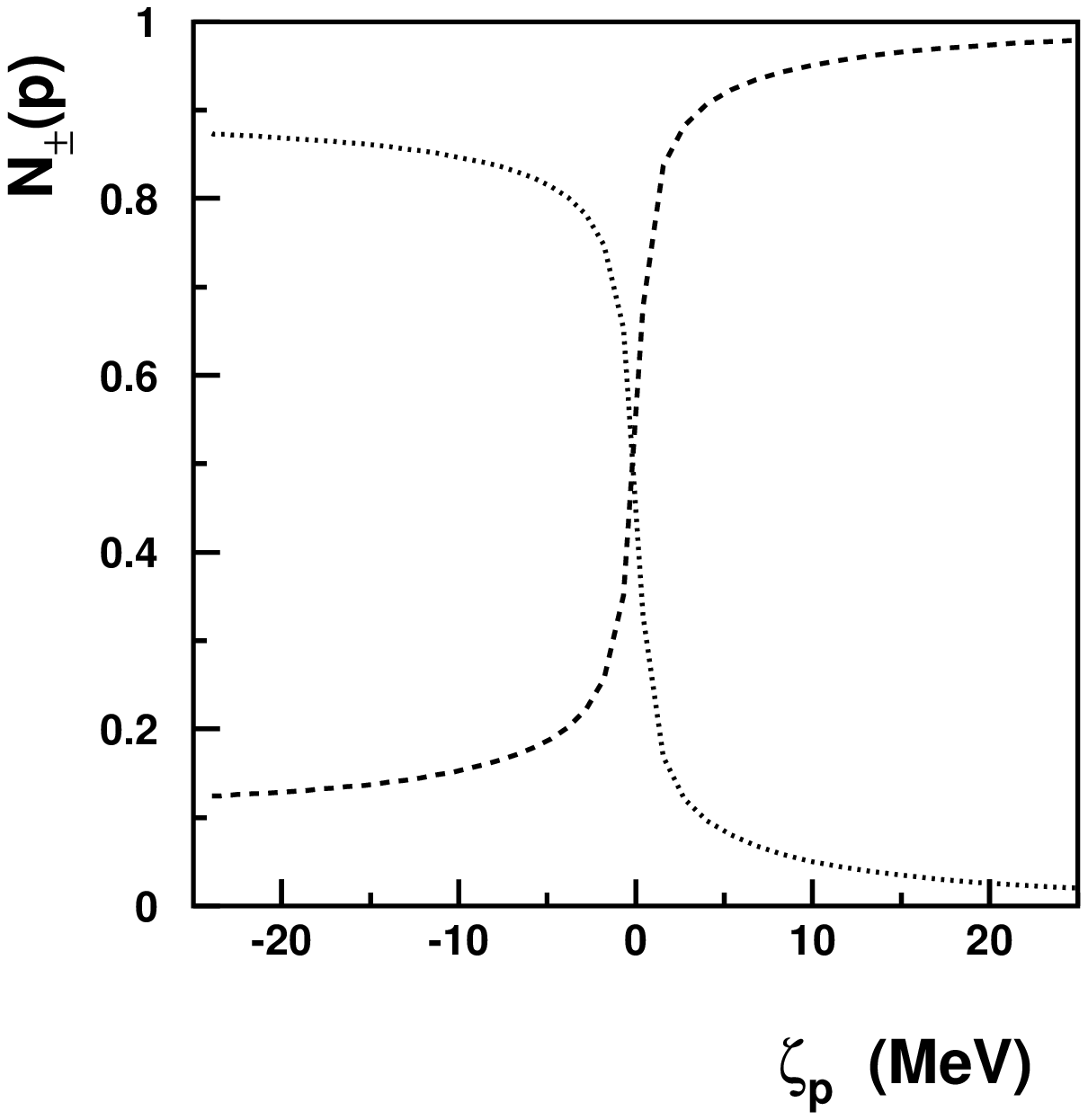,width=0.95\textwidth}
\caption{The weight of the peaks of the spectral function at positive and 
negative energies, for the same parameters as in Fig. \ref{spec17}.}
\label{wag17}
\end{minipage}
\end{figure}

The self-consistent solution has a similar  behavior around the critical
 temperature as the quasi-particle approximation:
at the critical temperature the T-matrix has a singularity at zero total 
momentum and energy, and for temperatures above $T_c$, the T-matrix is strongly
peaked for small values of energy and total momentum. 
This leads to a pseudogap structure in the spectral function
 (Fig. \ref{spec17}).
This pseudogap structure is again
similar to  the superfluid gap. Two peaks of the spectral 
function are located on both sides of the Fermi energy.
The dominant peak (with the larger weight) is the one whose energy 
is the closest to the energy of the quasi-particle pole.
Please note that for momenta
close to the Fermi momentum there is only one peak in the spectral function 
(the curve corresponding to $p=210$ MeV in Fig. \ref{spec17}).
In Fig. \ref{gap17} the dispersion of the two peaks of the spectral function is
presented. The dispersion of the maximum of the two peaks is very reminiscent 
of the superfluid
dispersion relations $E_k=\pm \sqrt{\zeta_k^2+\Delta(k)^2}$, where $\Delta(k)$ 
is 
the pseudogap. 
At the Fermi energy the separation between the two peaks should be equal 
to twice the value of the pseudogap. However, close to the Fermi energy only 
one, dominant peak of the spectral function survives.
Its width is small close to the Fermi energy, and the position of its
maximum is given by the 
energy of the quasi-particle pole $\zeta_p$.

It is difficult to 
devise a self-consistent approximate scheme, using a two-pole ansatz for the
spectral function. Usually such a procedure would require, when iterating
 Eq. (\ref{imsc}), the introduction of additional poles in the spectral 
function. A self-consistent approximate solution in the pseudogap region can 
be obtained only in the pairing approximation \cite{chica1}, i.e.
using a mean-field propagator on the right-hand side of Eq. (\ref{imsc}).
If additional assumptions are made for the T-matrix in the pseudogap region, a 
self-consistent approximation scheme with superfluid-like ansatz for
the spectral function can be obtained \cite{chica1}.
However, the self-consistent solution indicates (Fig. \ref{gap17}), that
 the pseudogap picture \cite{chica1} 
of single-particle excitations above $T_c$
is not entirely correct.
Moreover, the energy of the quasi-particle pole is different from the 
Hartree-Fock energy.
The imaginary 
part of the self-energy and its energy dependence are usually non-negligible.

In Fig. \ref{wag17} the integrated weights of the two peaks in the spectral 
function are presented.  The weight of the peak is obtained by integrating the
 spectral function on the same side of the Fermi energy.
Although qualitatively the weight of the 
two parts of the spectral function follows the superfluid result
$N_i(k)=\frac{1}{2}\Big( 1 \pm \frac{\zeta_k}{E_k}\Big)$, we could not fit 
the weight $N_i$ using this law.
This is not surprising since the two peaks are very broad, and,
moreover close to the Fermi energy where the effect of the pseudogap should be
 the clearest, the two peaks merge into one. The sum of the weights of the 
two peaks gives of course $1$.

\begin{figure}
\begin{minipage}[t]{0.48\linewidth}
\centering
\epsfig{file=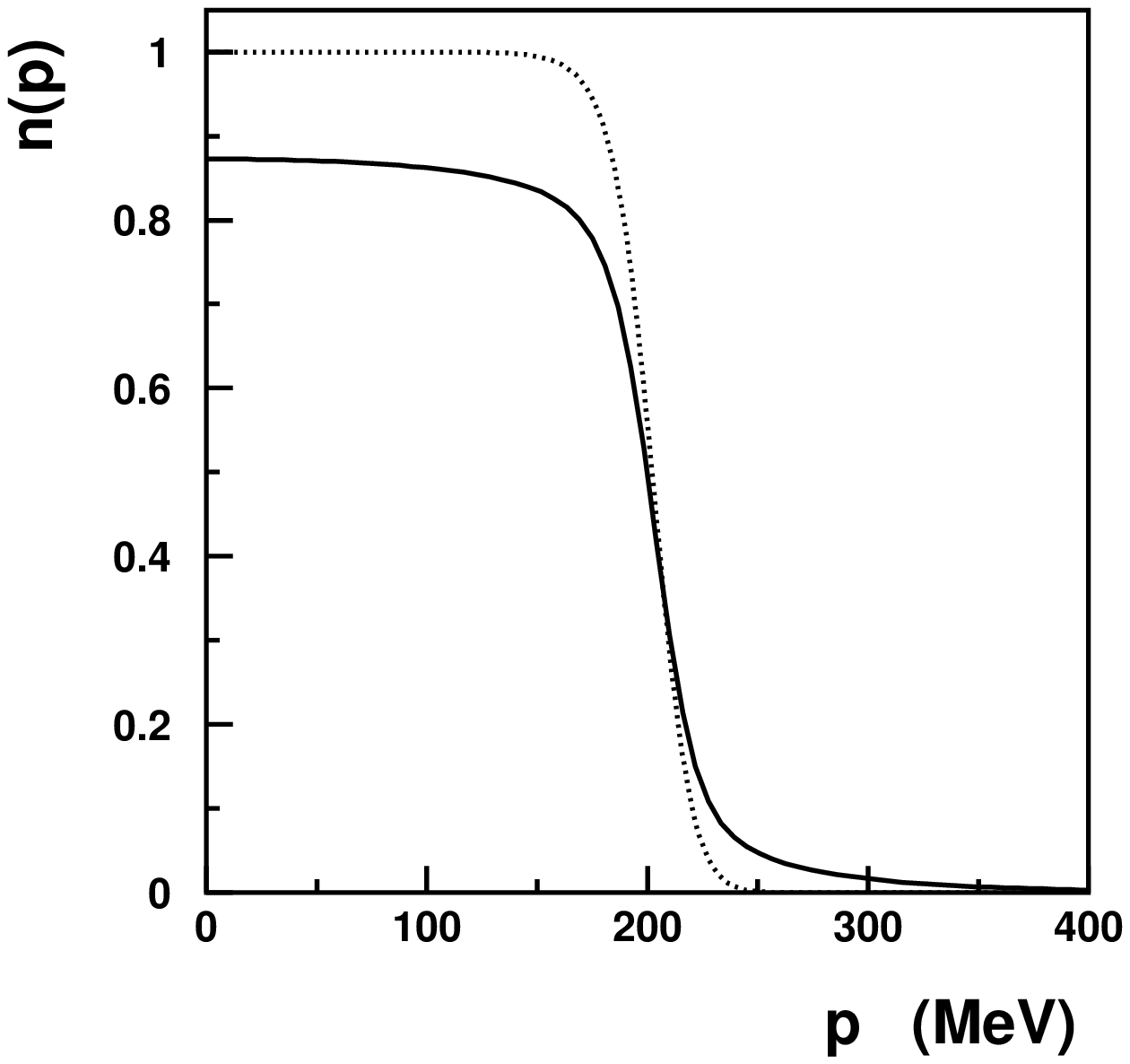,width=0.95\textwidth}
\caption{Nucleon momentum distribution obtained from the spectral 
function (Eq. \ref{mmdis}) (solid line)
 and using the quasi-particle pole approximation
 for the spectral function (dotted line),
 for the same parameters as in Fig. \ref{spec17}.}
 \label{fer17}
\end{minipage}
\hspace{.1in}
\begin{minipage}[t]{0.48\linewidth}
\centering
\epsfig{file=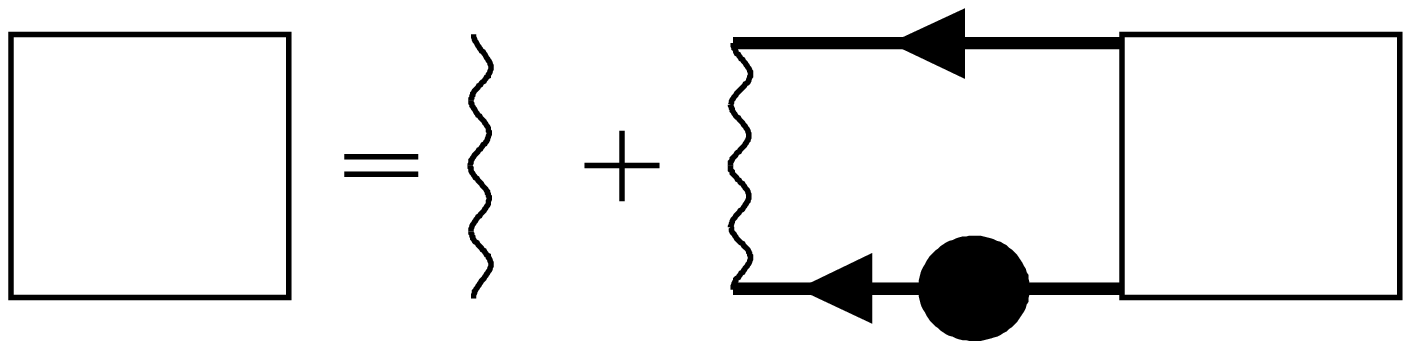,width=.99\textwidth}
\caption{Diagrammatic representation of the T-matrix equation in 
the superfluid phase. The thick line denotes the off-shell
 normal fermion propagator, the thick line with a dot denotes the full
fermion propagator including the anomalous self-energy, 
the wavy line denotes the interaction and the box stands for the T-matrix.}
\label{tsfig}
\end{minipage}
\end{figure}

The nucleon momentum distribution is presented in Fig. \ref{fer17}. 
The solid line was obtained using the self-consistently calculated spectral 
function
\begin{equation}
\label{mmdis}
n(p)=\int \frac{d \omega}{2 \pi} A(p,\omega) f(\omega) \ .
\end{equation}
The dotted line denotes the result of the quasi-particle approximation for 
the spectral function
\begin{equation}
n(p)=f(\zeta_p) \ ,
\end{equation}
where $\zeta_p$ denotes the position of the quasi-particle pole, i.e the 
solution of Eq. (\ref{sieq}).
Nucleon  scattering (nonzero imaginary part of the self-energy in $A$)
leads to a considerable broadening of the momentum distribution.
This is another indication that the quasi-particle approximation is
not reliable for nuclear matter around the critical temperature.



\section{Pairing and the T-matrix approximation}
\label{pairing}

In the superfluid phase the nuclear matter develops a superfluid energy gap
and non-zero anomalous fermion propagators. The usual BCS theory is limited 
to mean-field approximations to the self-energy and to the superfluid gap.
Formally T-matrix equations can be written for the normal propagators 
and anomalous propagators \cite{hau1,peder}. As before, T-matrix enters in the
 calculation  of the self-energy in place of the interaction potential.
But also
 a generalized T-matrix, between the anomalous propagators enters in the
calculation of the anomalous self-energy.

The resulting equations are, however, very difficult and no solution in this
 scheme has been presented. In the following we will present the simplest 
generalization of the T-matrix and BCS schemes, allowing for the calculation
 of the spectral function in the superfluid phase.
It is known that there is a link between the T-matrix and the BCS theory at the
critical temperature. The critical temperature in the BCS theory is related to
the appearance of a singularity in the T-matrix at zero energy \cite{thouless}.
It is a manifestation of more general relation between 
 the T-matrix fermion scattering and 
pairing, and the BCS gap equation. This relation was observed long time ago by
 Kadanoff and Martin \cite{km}. They proposed 
 to use one  mean-field fermion propagator 
and one BCS two-pole fermion propagator in the ladder
of the T-matrix diagram\footnote{We use the real time formalism also
in the superfluid phase \cite{rsf}, with energies measured with respect to 
the Fermi energy.}~:
\begin{eqnarray}
\label{teqkm}
<{\bf p}|T_{\alpha^{'}\beta^{'}\alpha\beta}
^\pm({\bf P},\omega)|{\bf p}^{'}> = V_{\alpha^{'}\beta^{'}\alpha\beta}
({\bf p},{\bf p}^{'}) \nonumber \\ + 
 \int\frac{d^3k}{(2 \pi)^3} \int \frac{d \omega^{'}}{2 \pi}
\int \frac{d \omega^{''}}{2 \pi} 
 V_{\alpha^{'}\beta^{'}\gamma\delta}
({\bf p},{\bf k}) \nonumber \\ 
\frac{A({\bf P}/2 +{\bf k},\omega^{'}-\omega^{''})
A_s({\bf P}/2 -{\bf k}, \omega^{''})\big(1-f(\omega^{'}-\omega^{''})-
f(\omega^{''})\big)}{\omega-\omega^{'}\pm i \epsilon} \nonumber \\
 <{\bf k}|T_{\gamma\delta\alpha\beta}^\pm({\bf P},\omega)
 |{\bf p}{'}> \ ,
\end{eqnarray}
where $A$ is the spectral function without anomalous self-energy, which we 
will call in the following normal fermion spectral function, and which takes 
the form 
\begin{equation}
A(p,\omega)=2 \pi \delta(\omega -\xi_p) \ ,
\end{equation}
in the quasi-particle  approximation, and
$A_s$ is the spectral function with anomalous self-energy, which we call
full fermion spectral function in the rest of the paper (Fig \ref{tsfig}). 
In the mean-field 
approximation it takes the usual BCS  form 
\begin{equation}
A_s({\bf p},\omega)=2 \pi \Bigg( \frac{E_{\bf k}+\xi_k}{2E_{\bf k}} 
\delta(\omega-E_{\bf k})
+\frac{E_{\bf k}-\xi_k}{2E_{\bf k}} \delta(\omega+E_{\bf k}) \Bigg) \ ,
\end{equation}
where
\begin{equation}
E_{\bf k}=\sqrt{\xi_k^2+\Delta({\bf k})^2} \ .
\end{equation}
Using (\ref{teqkm}) the gap equation can be written as 
\begin{equation}
\label{intgap}
\sum_{\alpha \beta}\int \frac{ d^3 p^{'}}{(2 \pi)^3}
<{\bf p}|{\rm Re} T_{\alpha^{'}\beta^{'}\alpha\beta}
^{\pm \ \ -1}({\bf P=0},\omega=0)|{\bf p}^{'}> \Delta_{\alpha \beta}
({\bf p^{'}}) = 0 \ ,
\end{equation}
the relation between $\Delta$ and $\Delta_{\alpha \beta}$ is given later.
The limit of the above equation when $\Delta=0$ in the T-matrix equation, 
gives the Thouless criterion for $T_c$. The meaning of Eq. (\ref{intgap})
is, however, more general. It states that the gap equation is equivalent to
requiring that the singularity of the T-matrix remains at zero energy and 
momentum for all temperatures below $T_c$.
The principle of conserving
 approximation, in the sense of Kadanoff and Baym \cite{kb}, requires 
the use of the normal fermion mean field propagator 
(without anomalous self-energy) on the right hand side of 
Eqs. (\ref{imsc}) and (\ref{hfsc}).

In the above equations we have assumed that the ground state is time-reversal
 invariant. Thus, the full fermion propagator $G_s$
 is still diagonal in spin-isospin
indices \cite{good,strbcs}.
 However, the full spectral function and the full  propagator
depend in general on the direction of the momentum. Also the T-matrix would 
then depend on the direction of the total momentum of the pair.

In principle the pairing approximation of Kadanoff and Martin can be used to 
calculate the spectral properties of the nuclear matter in the superfluid phase
using the quasi-particle approximation. However, as we have seen, the 
quasiparticle approximation 
becomes unreliable in the case of strong attractive interaction between 
fermions, with bound or almost bound states in vacuum.
As discussed in models of high $T_c$ superconductivity,
 the crossover region between the BCS pairing and the boson condensation,
 cannot be described using  mean-field
fermion propagators for the calculation of the T-matrix.
 The same is true in 
nuclear matter for with the Yamaguchi interaction in the T-matrix
 approximation, because of the pseudogap formation in the spectral function
(Sec. \ref{pseudosec}). At higher densities, in 
particular at the saturation density,
 the pairing gap is expected 
to be smaller and the BCS theory is correct. 
 Certainly calculations using more realistic interactions 
for several densities, including the modification of 
single-particle energies and finite quasi-particle width, would  
clarify at which densities, if any, the crossover-like behavior
 between the BCS pairing and 
the boson condensation occurs.
One way to generalize the Kadanoff-Martin approach was proposed by the Chicago 
group \cite{chica1,chica3}. From the observation that a pseudogap in 
the spectral function forms above $T_c$, the authors of Refs. 
\cite{chica1,chica3} proposed to use one mean-field 
and one full propagator in the T-matrix equation. The full propagator would 
include self-consistently the normal  self-energy in the T-matrix 
approximation (defined by (\ref{simfqp}) and (\ref{imqp})),
 and, below $T_c$, also the anomalous self-energy 
( defined by 
the superfluid gap (\ref{intgap})).  In  actual calculations, the authors of 
\cite{chica1,chica3} used an approximate BCS-like form also  the 
normal self-energy in the vicinity and below $T_c$. The spectral function
is then given in an analytical form by the superfluid gap and by a pseudogap,
defined by the singularity of the T-matrix for small energy and
small total momentum.
This approach represents an improvement over the quasi-particle
 approximation, because it takes into account the pseudogap formation 
in the calculation of the T-matrix. Also its a very nice illustration of the 
mechanism of the pseudogap formation and of the 
two-fermion excitations close to the
Fermi energy \cite{chica1}. The pseudogap, as well as the superfluid gap,
causes the formation of an energy gap in the two-fermion excitations
around the Fermi energy, very similar to the one postulated in \cite{d2}. 
 However, the self-consistent spectral function
 calculated in Sec. \ref{pseudosecsc} shows only partly a pseudogap.
The position of the subdominant peak in the spectral function 
does not follow the relation expected from a pseudogap energy dispersion 
relation. Also close to the Fermi energy only one peak in the spectral 
function is present. One could use the full spectral function instead of the 
pseudogap approximation for one of the propagators in the T-matrix equation. 
Thus, the correct dispersion of the peaks in the spectral function would be 
taken into account self-consistently. However, it turns out that the 
numerical cost of a calculation in the pairing approximation
 with one off-shell propagator and of a full
 self-consistent calculation is similar.
Also the use of the mean-field propagator in the calculation of the
self-energy (\ref{imqp}), as required in the pairing approximation,
 leads to a singularity in the self-energy at $T_c$. No such effect is 
present in the self-energy from the self-consistent calculation \cite{ja}.

\begin{figure}
\begin{minipage}[t]{0.48\linewidth}
\centering
\epsfig{file=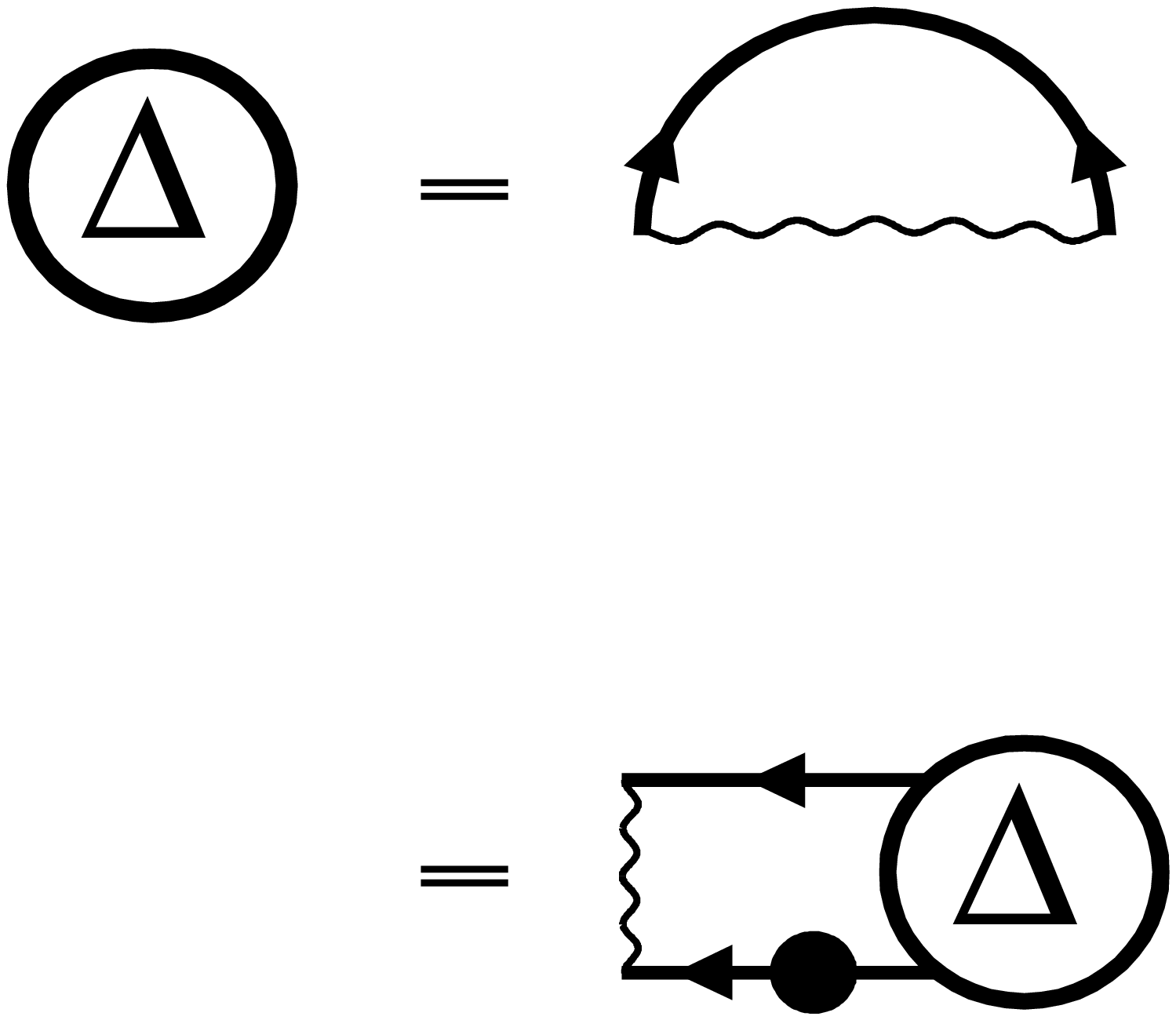,width=0.7\textwidth}
\caption{Diagrammatic representation of the gap equation (\ref{tengap})
 (first  line). The thick line with two arrows denotes the anomalous
 propagator and the  circle with the letter $\Delta$ denotes the pairing gap,
other symbols as in Fig. \ref{tsfig}. The equality in the second line,
is equivalent to Eqs. (\ref{intgap}), (\ref{gapzwyk}).}
 \label{dgapfig}
\end{minipage}
\hspace{.1in}
\begin{minipage}[t]{0.48\linewidth}
\centering
\epsfig{file=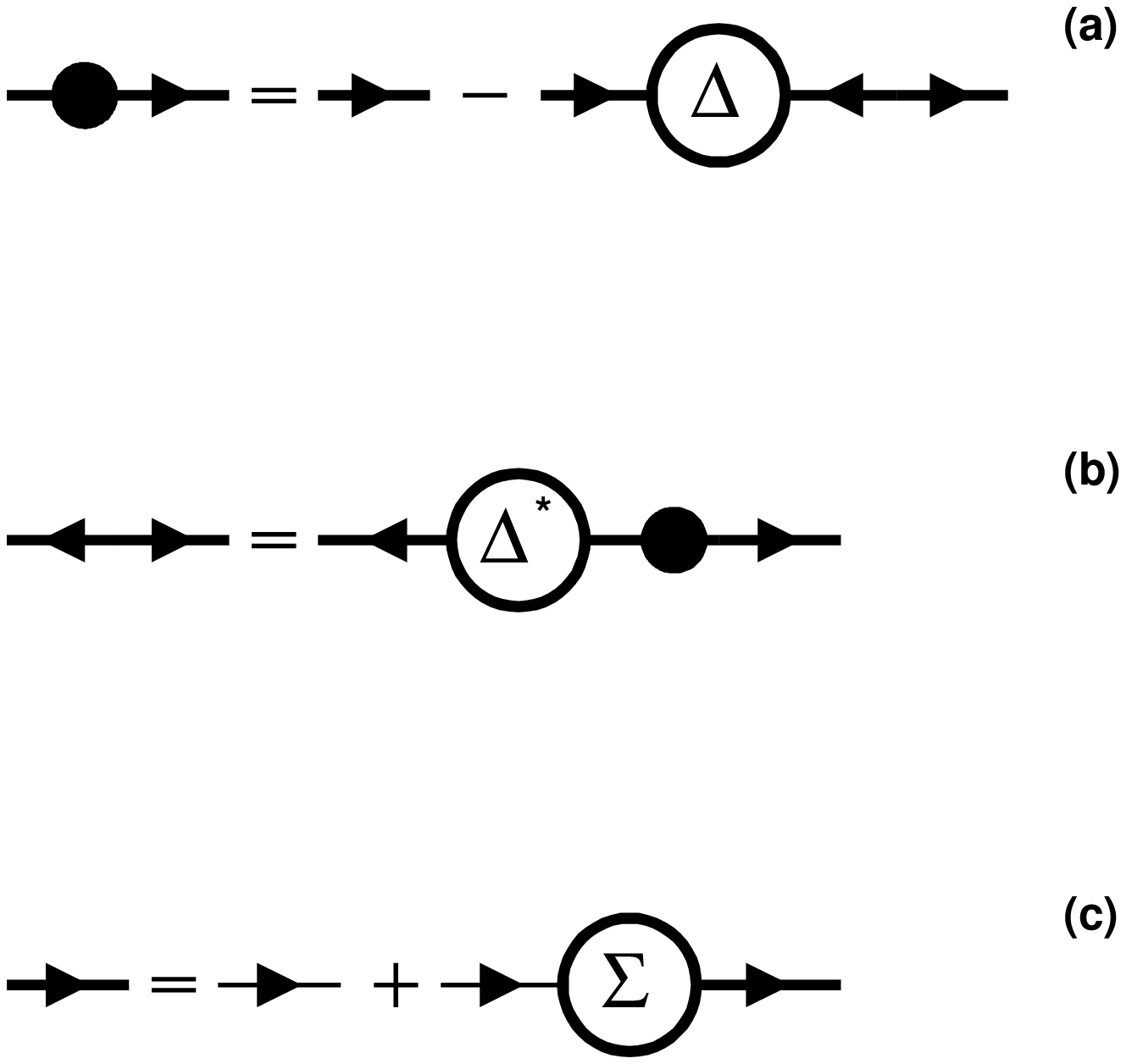,width=0.95\textwidth}
\caption{Diagrammatic representation of the BCS Dyson equation. 
{\bf (a)} represents the resummation of the anomalous self-energy in the
 nucleon propagator in the superfluid phase (Eq. \ref{strongBCS}),
{\bf (b)} represents the equation for the anomalous propagator 
(Eq. \ref{strongBCS2}) and {\bf (c)} represents the usual
 resummation of the self-energy (the circle with the  letter $\Sigma$).
Symbols as in Figs. \ref{tsfig}, \ref{dgapfig}.}
\label{dbcsfig}
\end{minipage}
\end{figure}

Therefore, we shall seek a generalization of the self-consistent
T-matrix equations in the
superfluid phase, including the off-shell fermion propagators in the T-matrix
equation and in the self-energy calculation. Since the T-matrix equation 
in the pairing approximation is related  to weak-coupling BCS equation 
\cite{km}, we will use a mean-field approximation for the anomalous 
self-energy.
Indeed the superfluid gap equation (Fig \ref{dgapfig})
\begin{equation}
\label{tengap}
\Delta_{  \alpha \beta}({\bf p}) = -\sum_{\alpha^{'}\beta^{'}}
\int \frac{d \omega}{ 2 \pi}
\int \frac{d^3 k}{(2 \pi)^3} V_{\alpha \beta \alpha^{'} \beta^{'}} 
\big(({\bf p}-{\bf k})/2,({\bf p}-{\bf k})/2 \big)
B_{\alpha^{'} \beta^{'}}({\bf k },\omega) f(\omega)  \ ,
\end{equation}
can be related to the T-matrix equation, using 
strong coupling BCS equations, with frequency independent gap 
parameter \cite{migdal,mahan} (Fig \ref{dbcsfig})
\begin{eqnarray}
\label{strongBCS}
G^{+}_s({\bf p},\omega)= 
G^{+}({\bf p},\omega)- \frac{1}{4}\sum_{\alpha \beta}
\Delta^{\alpha \beta}({\bf p})
F^{\dagger \ +}_{\alpha \beta}({\bf p},\omega)  \\
\label{strongBCS2}
F^{\dagger \ +}_{\alpha \beta}({\bf p},\omega)= G^{-}(-{\bf p}
,-\omega)G^{+}_s({\bf p},\omega)
 \Delta^{\dagger}_{\alpha \beta}({\bf p}) \ ,
\end{eqnarray}
where $G_s$ is the full fermion Green's function, including the 
anomalous self-energy ($G$ is reserved,
 from now on, to the normal Green's function 
including only the normal self-energy). 
The anomalous propagator spectral function is
\begin{equation}
B_{\alpha \beta}({\bf p},\omega)= -2 {\rm  Im}
 F^{+}_{\alpha \beta}({\bf p},\omega) 
\ .
\end{equation}
Using  (\ref{strongBCS2}) in the gap equation (\ref{tengap}),
and using the spectral representation of the propagators $G$ and $G_s$ one
obtains
\begin{eqnarray}
\label{gapzwyk}
 \Delta_{\alpha^{'}\beta^{'}}
({\bf p}) + 
 \int\frac{d^3k}{(2 \pi)^3} \int \frac{d \omega}{2 \pi}
\int \frac{d \omega^{'}}{2 \pi} 
 V_{\alpha^{'}\beta^{'}\gamma\delta}
({\bf p},{\bf k}) \nonumber \\ 
\frac{A({\bf P}/2 +{\bf k},\omega-\omega^{'})
A_s({\bf P}/2 -{\bf k}, \omega)\big(1-f(\omega-\omega^{'})-
f(\omega^{'})\big)}{\omega} \nonumber \\
 \Delta_{\gamma\delta}({\bf k}) = 0 \ .
\end{eqnarray}
Thus, the gap equation for the superfluid gap is equivalent to the
condition that $\Delta$ is the zero eigenvalue vector of the real part of the
inverse T-matrix (Eq. \ref{intgap}), i.e Eqs. (\ref{intgap}), (\ref{gapzwyk})
 and
(\ref{tengap}) are equivalent. However, here the T-matrix is given by 
the expression (\ref{teqkm}), with the spectral functions $A$ and $A_s$ being 
obtained from the  solution of (\ref{strongBCS}) 
($A(p,\omega)=-2{\rm Im}G^{+}(p,\omega)$, 
$A_s({\bf p},\omega)=-2{\rm Im}G^{+}_s
({\bf p}\omega)$).

The set of equations (\ref{teqkm}), (\ref{strongBCS}), and (\ref{tengap}),
defines a consistent approximation scheme for the nucleon self-energy 
below $T_c$. In this approach the  singularity of the T-matrix is always 
at zero energy and zero total momentum, the same is true in the BCS approach
\cite{km}. However here we include the contribution of the scattering 
and resonant  pairs to the self-energy, through Eq. (\ref{imsc}), 
as well as of 
the condensed pairs, through the anomalous self-energy.
In the region of intermediate 
coupling strength, between the weak BCS and the Bose 
condensation, the contribution of the T-matrix to the normal 
self-energy is important
\cite{hau1,pseudocm,chica1,chica3}, and leads to a pseudogap formation.

There is one more technical obstacle before using this scheme in actual 
calculations. In the superfluid phase the full propagator 
$G_s({\bf p},\omega)$ depends on the
direction of the momentum ${\bf p}$, and the 
T-matrix depends on the direction of the total momentum. This dependence 
introduces tremendous difficulties in the numerical solution of the T-matrix 
(\ref{teqkm}) and the BCS Dyson equation (\ref{strongBCS}).
Also in the T-matrix calculation we have used partial-wave expansion, 
enabled by the angle averaging of the two-fermion propagator in the 
ladder diagrams. On the other hand, the solution of the 
superfluid gap equation, which is
equivalent  to the  calculation
T-matrix equation at zero total momentum, requires no angular averaging of the 
two propagators in (\ref{strongBCS2}). 
In the BCS theory an approximate decoupling of different angular
 momentum gaps is obtained \cite{anmo}.
In that case the gap equation in each partial wave approximately decouple, 
i.e. we can use a different propagator $G_s$ in  each of the
partial wave gap equations. 
In nuclear physics the situation is more complicated due to the coupled
partial waves in the T-matrix. If we include the coupled partial waves, 
we cannot use the partial wave decoupling of the gap equations
(e.g. $^3S_1-^3D_1$).
Also in the nuclear matter we can in general have two superfluid gaps
for the same partial wave but with different total angular momentum and isospin
(e.g. $^1S_0$ and $^3S_1$). Moreover, we want to use the angular averaged
two-nucleon propagator in the T-matrix equation. Thus, the corresponding
gap equations in different channels will be coupled through the dependence
of the full fermion propagator $G_s$ on the anomalous self-energy.
In practice only the gap corresponding to one channel  would be non-zero,
the one with the strongest pairing at a given temperature and density. 
The T-matrix in other channels would have the singularity pushed away
 from the Fermi energy due to the energy gap originating form the
paired channel. Of course this does not exclude phase transitions between
pairing in different channels, when changing the temperature or density.


In order to simplify the description of the full fermion propagator we shall
 use an angle averaged anomalous self-energy in (\ref{strongBCS}). 
This approximation  simplifies also the gap equation \cite{strbcs}, in 
particular the superfluid gap for non-zero angular momentum can be taken as 
real and independent of the projection of the angular momentum \cite{bawe}.
Of course such an assumption for the superfluid gap is only an approximation,
and in some cases the true ground state of the nuclear matter may be described
by  more complicated angle dependent anomalous propagators. The example
of the liquid Helium teaches us that the structure of the anisotropic
 superfluid order parameter can be very rich \cite{anisotropic}. 
In particular the real order parameter would not lead to the state of lowest 
energy. However, it is sufficient for 
consistent and stable T-matrix calculation of 
the nuclear matter. It would be extremely difficult to 
implement the most general
angle dependent superfluid gap and full fermion propagator in the
T-matrix calculation. 

As a result the self-consistent T-matrix can be decomposed in partial waves
\begin{eqnarray}
\label{tsf}
<{ p}|T^{(JST)  \ \pm}_{ll^{'}}
({ P},\omega)|{ p}^{'}> = V^{(JST)}_{ll^{'}}(p,p^{'}) \nonumber \\
+  \sum_{l^{''}}
 \int\frac{d^3k}{(2 \pi)^3} \int \frac{d \omega^{'}}{2 \pi}
\int \frac{d \omega^{''}}{2 \pi} 
 V^{(JST)}_{ll^{''}}
({ p},{ k}) \nonumber \\ 
\langle \frac{A(|{\bf P}/2 +{\bf p}|,\omega^{'}-\omega^{''})
A_s(|{\bf P}/2 -{\bf p}|, \omega^{''})\big(1-f(\omega^{'}-\omega^{''})-
f(\omega^{''})\big)}{\omega-\omega^{'}\pm i \epsilon} \rangle_{\Omega}
 \nonumber \\
 <{ k}|T^{(JST)  \ \pm}_{l^{''}l^{'}}({ P},\omega)
 |{ p}{'}> \ .
\end{eqnarray}
The full fermion Green's function can be written as
\begin{equation}
G^{+}_s(p,\omega)=\frac{1}{G^{+}(p,\omega)^{-1}+  \Delta^2(p) G^{-}(p,-\omega)}
\ ,
\end{equation}
where the normal fermion propagator is given by the standard expression 
 and $\Delta(p)$ is the angle averaged total energy gap. 
 The  self-energy $\Sigma$ is given by Eqs. (\ref{hfsc}) and
(\ref{imsc}), with the normal spectral function $A$ on the right hand side.
The set of gap equations in (JST) channels takes the form
\begin{eqnarray}
\label{gapsc}
\Delta_{l}^{(JST)}({ p}) = -\sum_{l^{'}} \int \frac{d \omega}{2 \pi}
\int \frac{d \omega^{'}}{ 2 \pi} \int \frac{k^2 dk}{(2 \pi)^3}
V_{ll^{'}}^{(JST)}(p,k) \nonumber \\
\frac{A(k,\omega-\omega^{'})A_s(k,\omega^{'})\big(1-f(\omega-\omega^{`})
-f(\omega^{'})\big)}{\omega} \Delta^{(JST)}_{l^{'}}(k) \ .
\end{eqnarray}
The existence of a non-zero solution of the above equation in a particular 
channel is equivalent to the presence of a singularity in the T-matrix in 
that channel at zero total momentum and zero energy.
The total, angle averaged
 energy gap is given by an
 incoherent sum of contributions in different channels
\begin{eqnarray}
\Delta^2(p)&=&\frac{1}{8 \pi} \sum_{(JST)l} (2T+1)(2J+1)
\Delta_{l}^{(JST)}(p)^2 \nonumber
 \\
&=& \frac{1}{8 \pi} (2T+1)(2J+1)\sum_{l} \Delta_{l}^{(JST)_s}(p)^2 \ ,
\end{eqnarray}
where in the last equality we have used the fact that only one channel
$(JST)_s$ gives nonzero gap. It is the channel which gives the larger value 
of $\Delta(p)$.
One can notice that using mean-field spectral function in (\ref{gapsc}),
one recovers the  BCS gap equations in (JST) channels from \cite{strbcs}.
The final equation of the calculation scheme is the expression for the 
nuclear density
\begin{equation}
\label{rhosf}
\rho=4 \int \frac{d \omega}{2 \pi}\int \frac{d^3 p}{(2 \pi)^3}A_s(p,\omega)
f(\omega) \ .
\end{equation}
It is instructive to write explicitly the 
full fermion spectral function in terms
of the energy gap and the self-energy
\begin{eqnarray}
\label{As}
A_s(p,\omega)&=&-2\Bigg(\big(\omega+\xi_p+{\rm Re}\Sigma^{+}(p,-\omega)\big)^2
{\rm Im}\Sigma^{+}(p,\omega) \nonumber \\& &
+{\rm Im}\Sigma^{+}(p,-\omega) \Delta^2(p) 
+ \big( {\rm Im}\Sigma^{+}(p,-\omega)\big)^2 {\rm Im} \Sigma(p,\omega)\Bigg)/
\nonumber \\& &
\Bigg(\Big(\big(\omega-\xi_p-{\rm Re}\Sigma^{+}(p,\omega)\big)
\big(\omega + \xi_p +{\rm Re}\Sigma^{+}(p,-\omega)\big) \nonumber \\ & &-
{\rm Im}\Sigma^{+}(p,\omega){\rm Im}\Sigma^{+}(p,-\omega)-\Delta^2(p)\Big)^2
\nonumber \\ & &
+\Big( {\rm Im}\Sigma^{+}(p,\omega)\big(\omega+\xi_p+{\rm Re}\Sigma^{+}(p,
-\omega)\big) \nonumber \\ & &+ {\rm Im}\Sigma^{+}(p,-\omega)\big(\omega-\xi_p-{\rm Re}
\Sigma^{+}(p,\omega)\big)\Big)^2 \Bigg) \ .
\end{eqnarray}
It is clear that in order to find the spectral function for energy $\omega$,
one has to know the self energy for energy $\omega$ and $-\omega$.
Therefore, the numerical solution must be performed in a symmetrical 
interval around the Fermi energy ($\omega=0$).

\section{Nuclear matter in the superfluid phase}
\label{super}

\begin{figure}
\begin{minipage}[t]{0.48\linewidth}
\centering
\epsfig{file=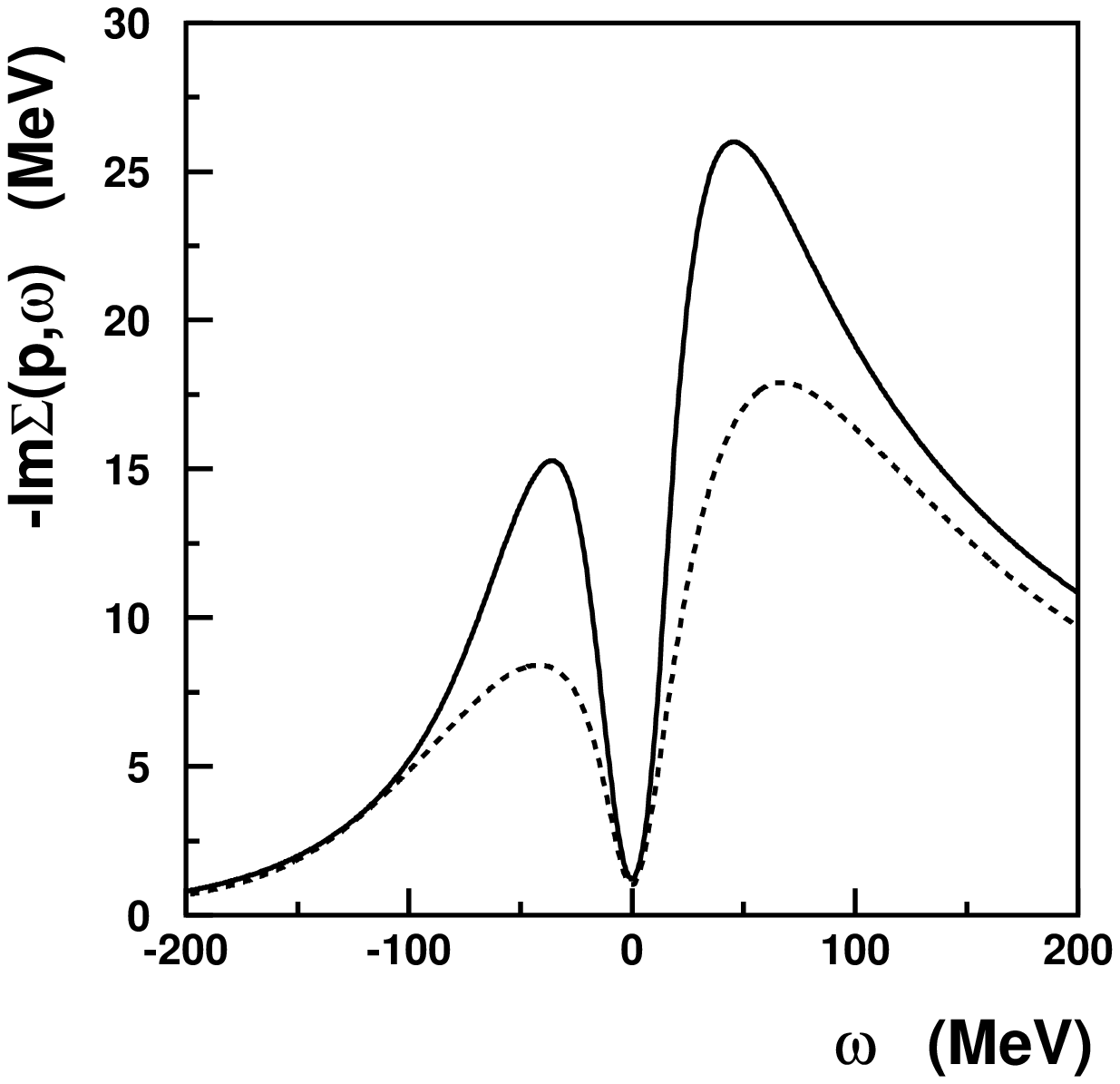,width=0.95\textwidth}
\caption{The imaginary part of the self-energy, calculated in the 
self-consistent T-matrix approximation as function of the energy,
for the same parameters as in Fig. \ref{spec17}. 
The solid and dashed lines denote the results at $p=0$ and $p=175$ MeV 
respectively.} 
\label{gam17}
\end{minipage}
\hspace{.1in}
\begin{minipage}[t]{0.48\linewidth}
\centering
\epsfig{file=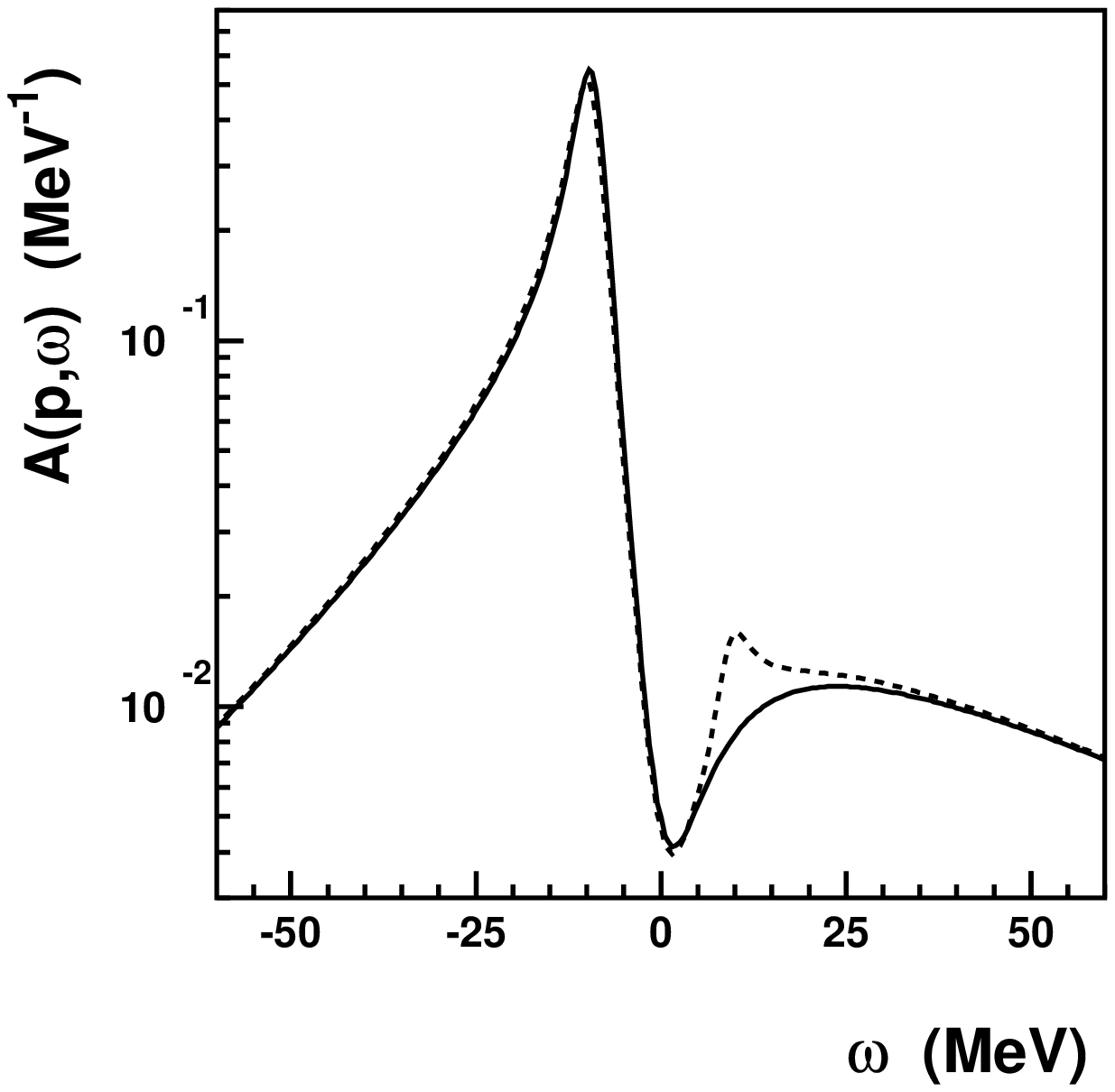,width=0.95\textwidth}
\caption{The spectral function $A(p,\omega)$ 
 (solid line) and the spectral function
with anomalous self-energy $A_s(p,\omega)$
(dashed line) as function of energy at $p=140$~MeV.
The results were obtained  at $T=1.4$MeV, and $\rho=.45\rho_0$,
corresponding to a superfluid gap of $\Delta(0)=5.1$~MeV.}
\label{spec14a}
\end{minipage}
\end{figure}

\begin{figure}
\begin{minipage}[t]{0.48\linewidth}
\centering
\epsfig{file=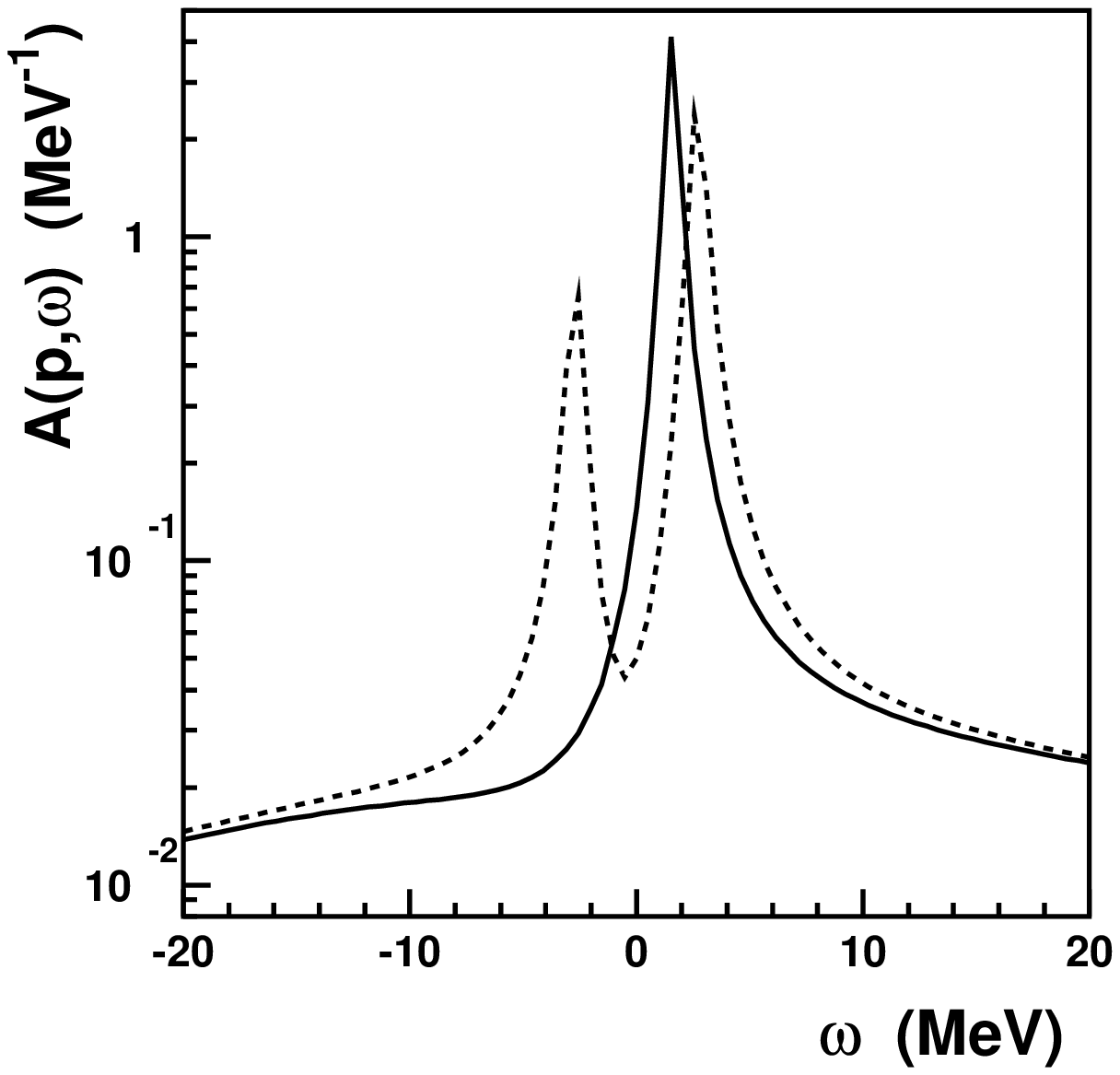,width=0.95\textwidth}
\caption{Same as in Fig. \ref{spec14a} but for $p=210$~MeV.}
\label{spec14b}
\end{minipage}
\hspace{.1in}
\begin{minipage}[t]{0.48\linewidth}
\centering
\epsfig{file=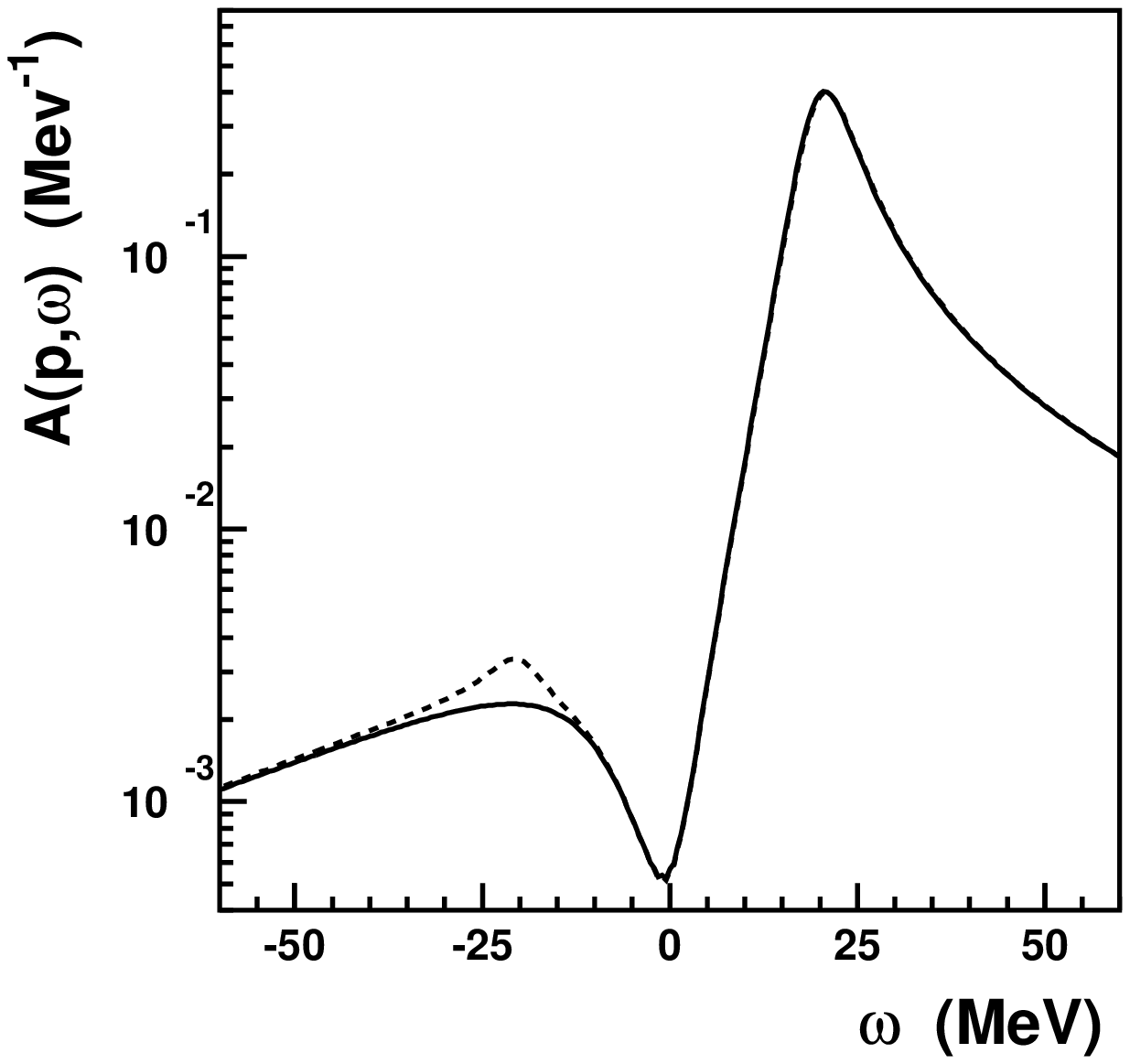,width=0.95\textwidth}
\caption{Same as in Fig. \ref{spec14a} but for $p=280$~MeV.}
\label{spec14c}
\end{minipage}
\end{figure}

We  solve numerically the set of equation (\ref{tsf}), 
(\ref{imsc}),  (\ref{As}), (\ref{gapsc}), with the constraint (\ref{rhosf})
for the same interaction as used in Sec. \ref{pseudosec}.
Eqs. (\ref{rhosf}) and (\ref{gapsc}) are solved simultaneously for the
superfluid gap $\Delta(k)$ and for the chemical potential  $\mu$.
Below $T_c$ the gap equation has a nontrivial solution, corresponding to the
superfluid phase. 
For the case of the Yamaguchi interaction,
 the pairing occurs always in the $^3S_1$ channel.
The T-matrix has a singularity at zero momentum and zero energy in this 
channel, for all temperatures below $T_c$. The T-matrix in the $^1S_0$ 
channel has no  singularity but is  strongly peaked near the Fermi energy.
For more realistic interaction we expect also pairing in the $1^S_0$ channel 
and pairing with higher angular momentum at larger densities.
The calculation gives two spectral functions, the normal spectral function
$A(p,\omega)$, without the anomalous self-energy, 
and the full spectral function
$A_{s}(p,\omega)$, obtained from the BCS Dyson equation (\ref{strongBCS}). 
We have observed that due to strong attractive interaction, and due to the 
appearance of the singularity in the T-matrix, a BCS-like two peak structure 
appears in the spectral function $A$ in the pseudogap region. The same is true
below $T_c$ (Figs. \ref{spec14a} and \ref{spec14c}). 
The spectral function without anomalous self-energy is very similar as above 
$T_c$ (Fig. \ref{spec17}). Indeed the imaginary part of the self-energy
(Fig. \ref{gam17})
is almost indistinguishable  for the two temperatures,
 except near the Fermi energy\footnote{As expected 
the scattering width at the Fermi 
energy decreases with the temperature.}.
In the full spectral
 function another mechanism is also responsible for the double peak structure.
  For all momenta the full spectral function
 $A_s$
has two peaks situated on both sides of the Fermi energy.
Close to the Fermi energy, the superfluid
 gap in the spectral function is clearly visible (Fig \ref{spec14b}).
Far from the Fermi energy the modifications of the dominant peak are 
negligible. 
The subdominant peak of the spectral function is only slightly modified
by the anomalous self-energy. This is what is expected form the usual 
BCS theory, since the weight of the subdominant superfluid peak in the
spectral function $\frac{1}{2}\Big(1-\frac{|\xi_k|}{E_k}\Big)$ is becoming
small far from the Fermi energy. 

\begin{figure}
\begin{minipage}[t]{0.48\linewidth}
\centering
\epsfig{file=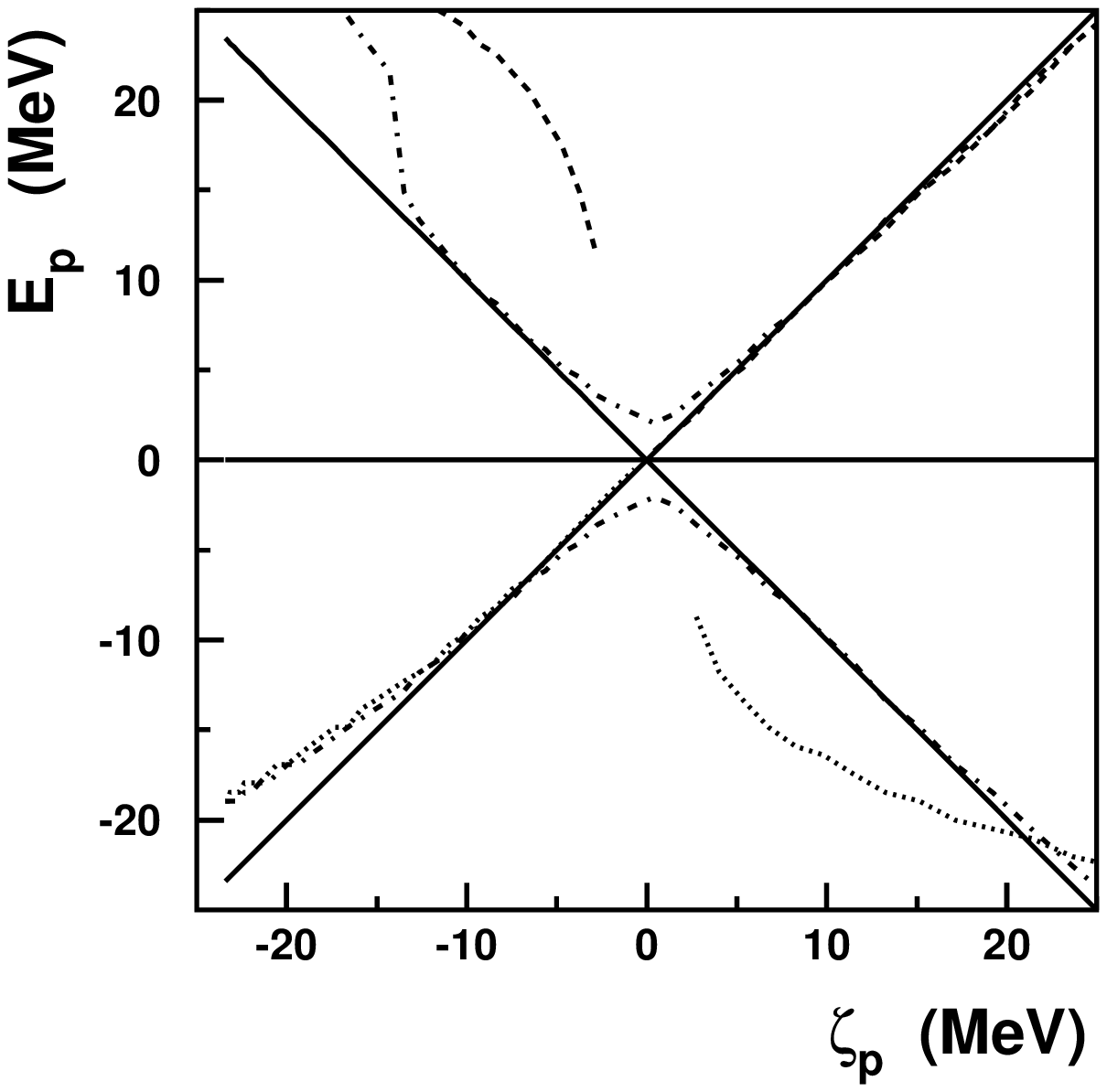,width=0.95\textwidth}
\caption{The positions of the two peaks of the normal spectral function
(dotted and dotted lines) and of the full spectral function
(dash-dotted  lines)
as function of the energy of the  quasi-particle pole,
 for the same parameters as in Fig. \ref{spec14a}.
 The solid lines represent the two asymptotic branches in the 
BCS solution $E_p=\pm \zeta_p$ and the Fermi energy $E_p=0$.}
\label{gap14}
\end{minipage}
\hspace{.1in}
\begin{minipage}[t]{0.48\linewidth}
\centering
\epsfig{file=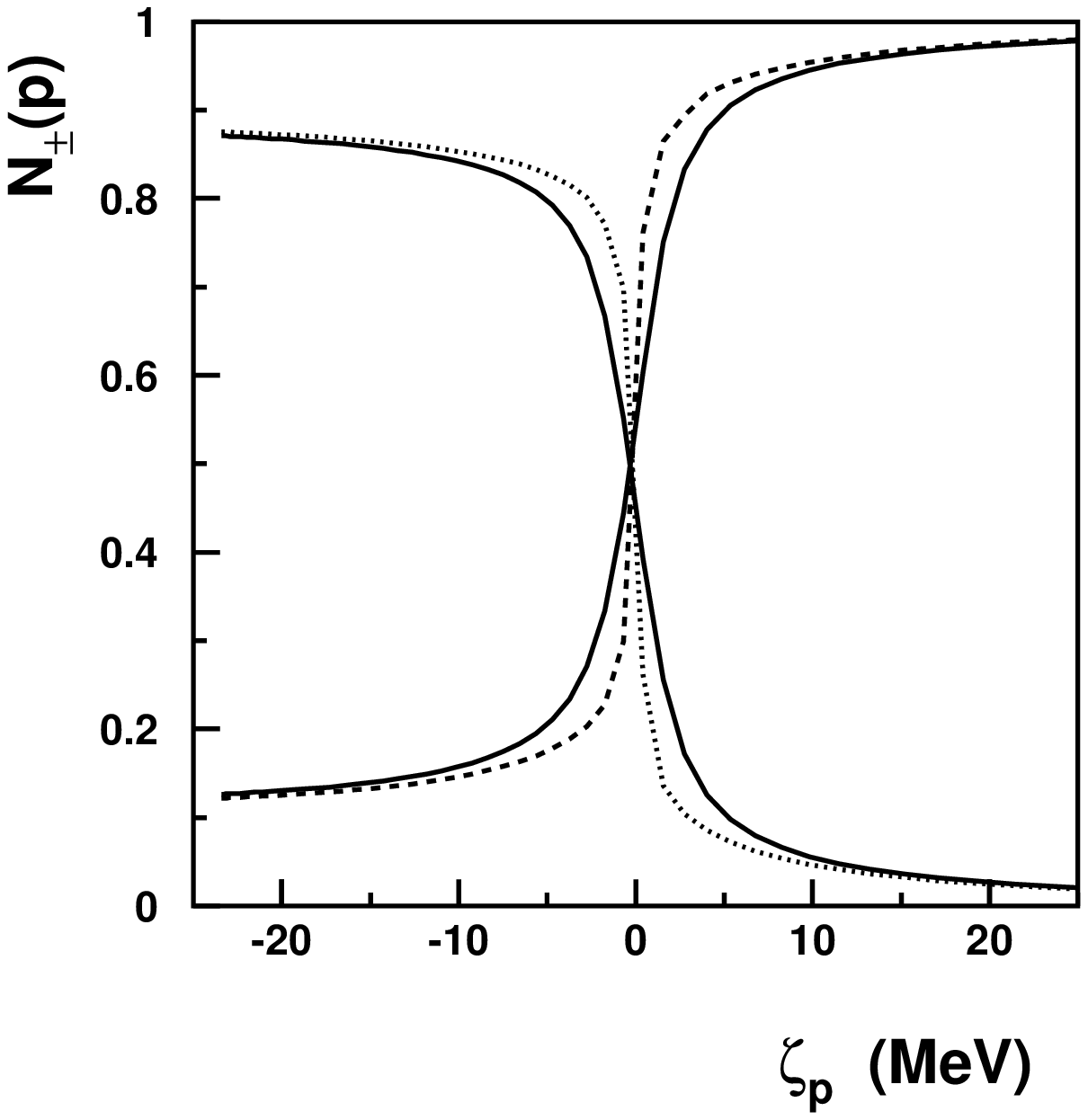,width=0.95\textwidth}
\caption{The weight of the peaks of the normal 
spectral function at positive and 
negative energies (dashed and dotted lines) and of the peaks
of the spectral function
 with anomalous self-energy (solid lines),
 for the same parameters as in Fig. \ref{spec14a}.}
\label{wag14}
\end{minipage}
\end{figure}

The dispersion of the maximum of the peaks of the spectral functions $A$
and $A_s$ illustrates the formation of the superfluid gap (Fig. \ref{gap14}).
On the abscissa we plot the energy of the quasi-particle pole in the normal
 fermion Green's function $G$. This energy coincides approximately with 
the maximum of the
dominant peak in the spectral function $A$. This is especially well fulfilled 
in the vicinity of the Fermi energy. The subdominant peak in the spectral 
function $A$ is visible for energies further away from the Fermi energy, 
also its energy does not follow the branch $E_p=-\zeta_p$. Like for the case 
above $T_c$, the pseudogap in the spectral function is only  qualitatively
similar to a superfluid gap. Close to the Fermi energy, the spectral function
$A$ is similar as in the normal Fermi liquid. 
On the other hand,
the positions of the peaks of the full spectral function $A_s$  follow
the dispersion relation expected for a superfluid matter, especially close to
the Fermi energy. An energy gap is clearly visible for $\zeta_p\simeq0$.
Away from the Fermi energy, the positions of the two peaks approach
$E_p=\zeta_p$ for the dominant peak and $E_p=-\zeta_p$ for the subdominant peak
in the full fermion spectral function. For $|\zeta_p|>10$~MeV, the positions
 of the  peaks do not follow the BCS energy branches any more.
 However, one can 
observe that the dominant peaks in the normal fermion spectral function and 
in the full fermion spectral function are very close away from the Fermi 
energy, as expected.

\begin{figure}
\begin{minipage}[t]{0.48\linewidth}
\centering
\epsfig{file=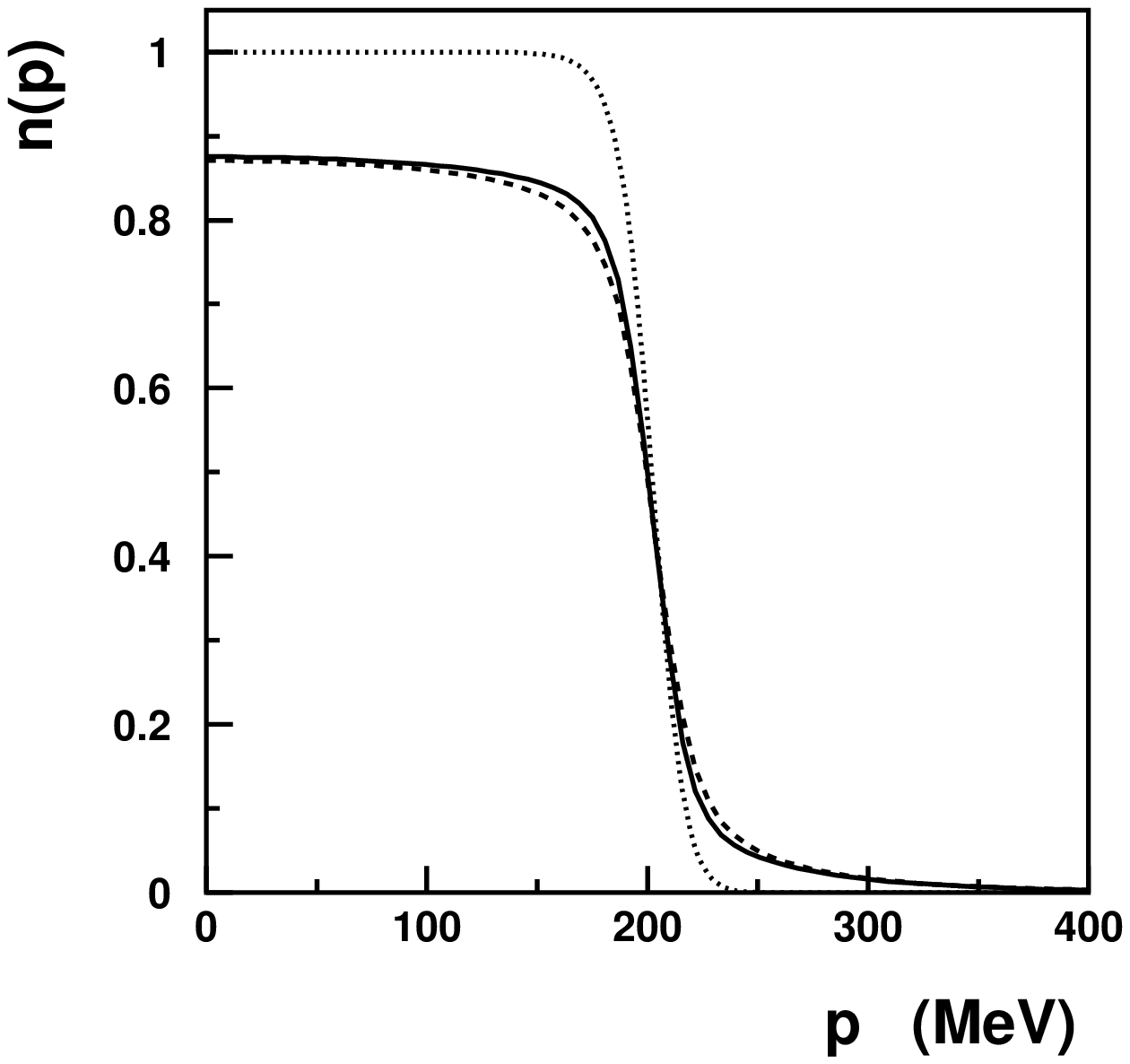,width=0.95\textwidth}
\caption{Nucleon momentum distribution obtained from the normal
spectral function (solid line), from the full spectral
function (dashed line), and using the quasi-particle pole approximation
for the spectral function (dotted line), 
for the same parameters as in Fig. \ref{spec14a}.}
 \label{fer14}
\end{minipage}
\hspace{.1in}
\begin{minipage}[t]{0.48\linewidth}
\centering
\epsfig{file=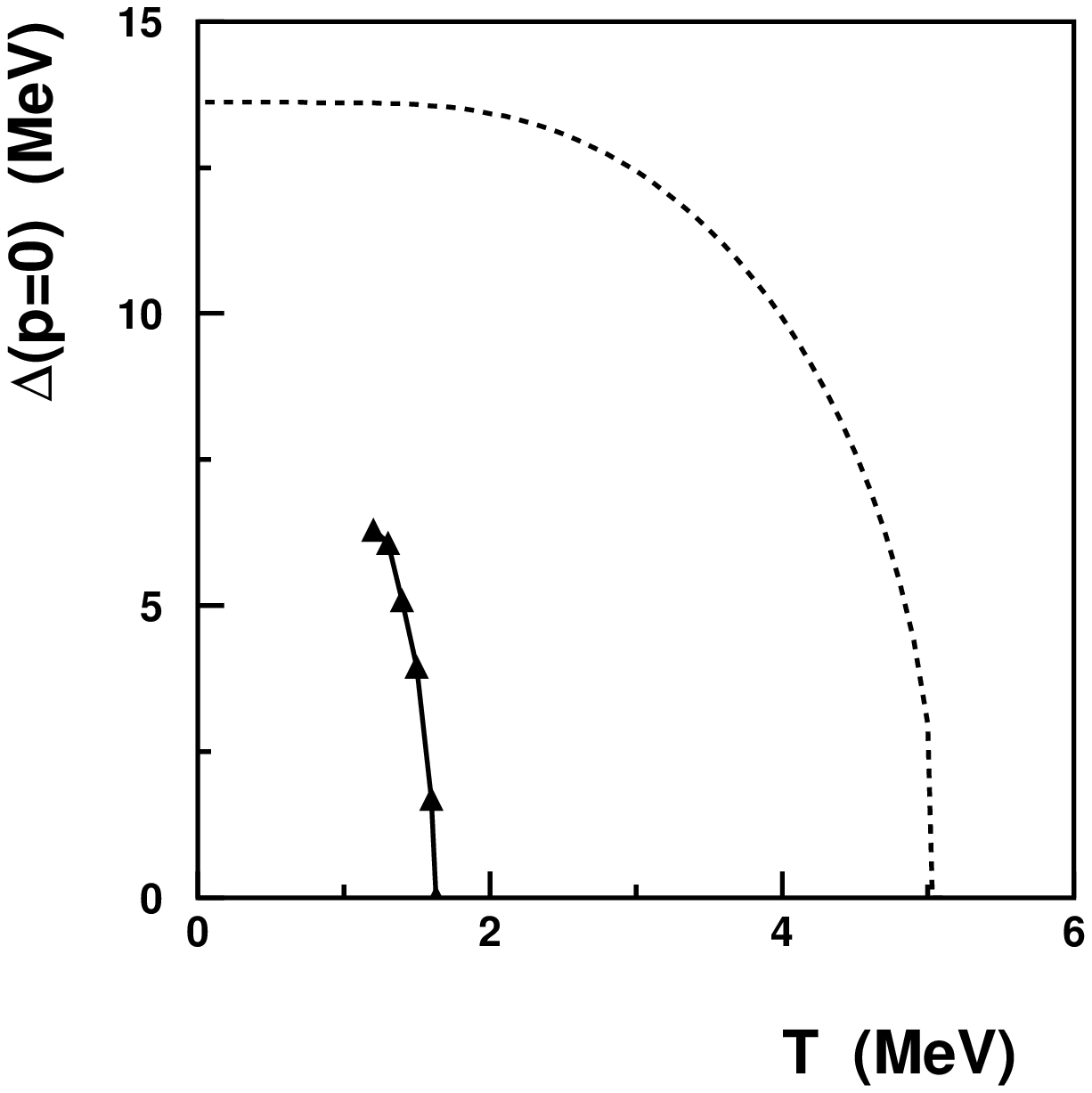,width=0.95\textwidth}
\caption{The superfluid energy gap as function of  temperature for
the BCS theory (dashed line) and for the self-consistent calculation
(triangles), for $\rho=.45\rho_0$.}
\label{szcz}
\end{minipage}
\end{figure}

\begin{figure}
\begin{minipage}[t]{0.48\linewidth}
\centering
\epsfig{file=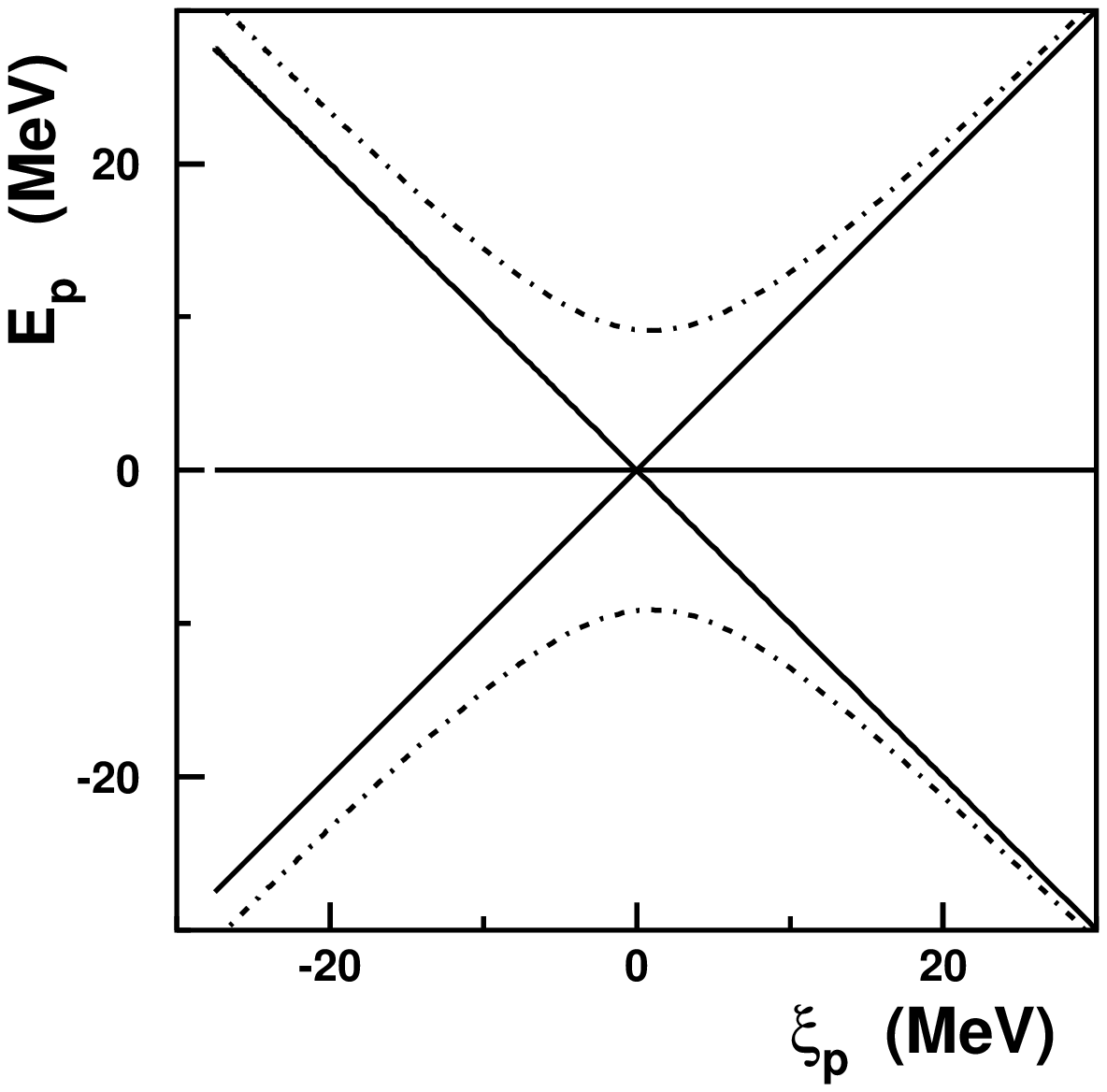,width=0.95\textwidth}
\caption{BCS quasiparticles energies 
 as function of the
Hartree-Fock single-particle energy, for $T=1.4$~MeV and $\rho=.45\rho_0$.}
\label{gapbcs}
\end{minipage}
\hspace{.1in}
\begin{minipage}[t]{0.48\linewidth}
\centering
\epsfig{file=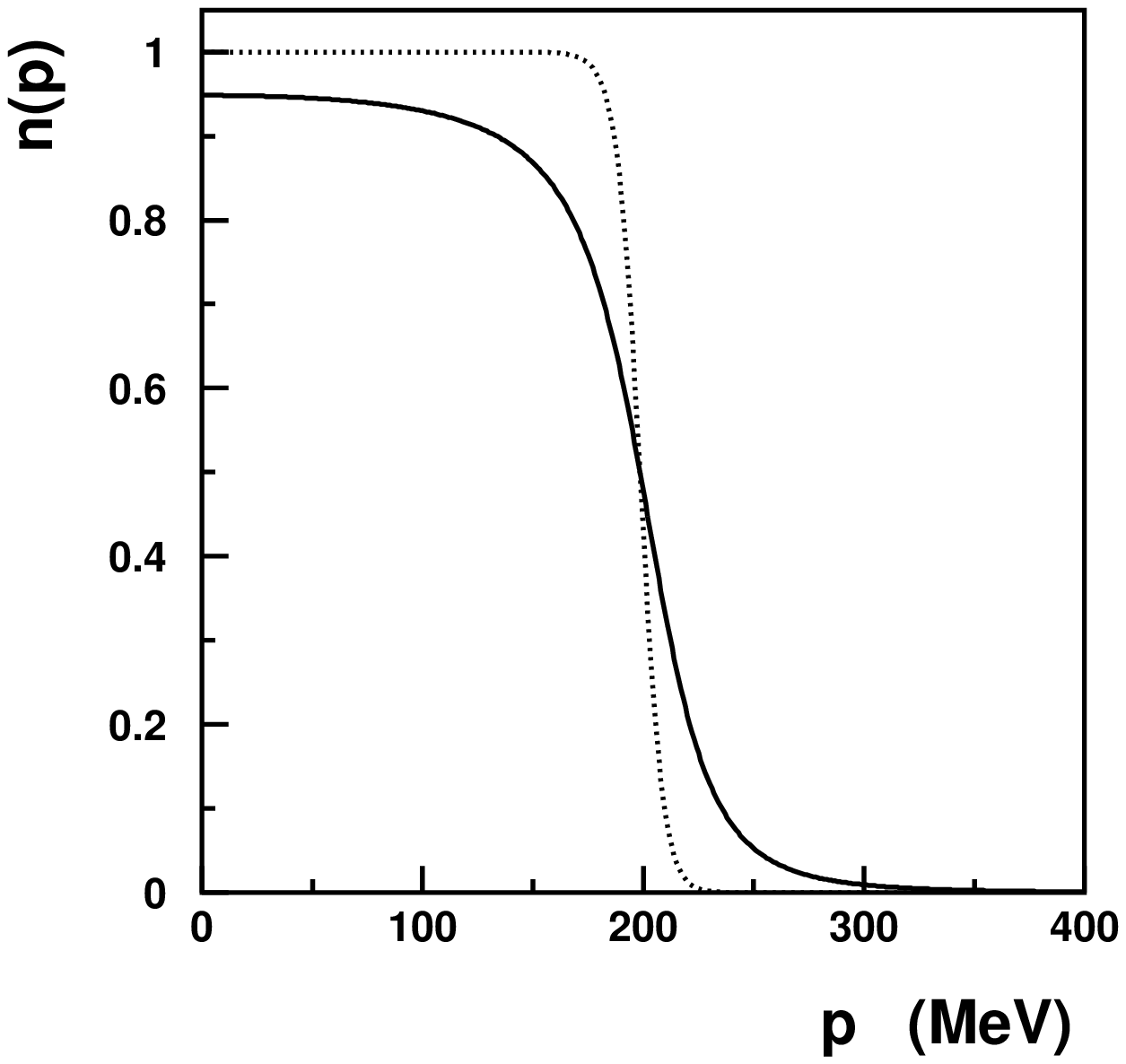,width=0.95\textwidth}
\caption{Nucleon momentum distribution for the Hartree-Fock 
single-particles energies (dotted line) and the BCS nucleon distribution 
(solid line), for the same parameters as in Fig. \ref{gapbcs}.}
\label{ferbcs}
\end{minipage}
\end{figure}

In Fig. \ref{wag14} are presented the weights of the peaks in the spectral 
functions $A$ and $A_s$ as function of the energy of the quasi-particle pole 
of 
the spectral function $A$. The peaks in the two spectral functions have 
qualitatively similar behavior.
 However,  the weight of the dominant (subdominant)
peak is smaller (larger) in the full spectral function, than in  the 
normal spectral function. It means that part of
 the weight of the spectral function is shifted from the dominant to the
subdominant peak, as expected from the presence of an additional mechanism
of peak doubling in the superfluid phase.

The fermion momentum distribution is plotted in Fig. \ref{fer14}.
The finite imaginary part of the spectral function modifies
 strongly the  momentum distribution. The momentum distributions obtained from 
the spectral functions $A$ and $A_s$ are much broader then the one obtained
taking the quasi-particle pole approximation for 
the spectral function. As expected,
the nucleon scattering produces a high momentum tail in the distributions.
The effect of the superfluid gap can be seen in the difference between the 
solid and dashed lines in Fig. \ref{fer14}. As expected \cite{migdal,mahan}
the full spectral function $A_s$ gives broader momentum distributions. 
The BCS broadening with small energy gap results in the smearing of the
 momentum distribution in an interval $2\Delta(p_F)$ around the Fermi energy.
The density given by the quasi-particle approximation to the spectral 
function $A$ is $\sim1.01$ times the actual fermion density given by 
(\ref{rhosf}). The densities
 calculated from the spectral function $A$ and $A_s$
are almost the same.

In Fig. \ref{szcz} we plot the temperature dependence of the superfluid gap
$\Delta(0)$ as function of the temperature. Below $T_c=1.63$ MeV
 the superfluid gap 
rapidly sets in. We could not perform calculations below $T=1.2$ MeV,
due to limited numerical resolution in energy. However, a saturation in the
dependence of the superfluid gap can already be observed at the lowest 
 temperatures studied. The dashed line denotes the result of the BCS theory.
The critical temperature for the  BCS theory is the same as in the 
quasi-particle approximation, $T_c=5.03$MeV. 
The scattering of nucleons and the pseudogap formation above $T_c$ 
make the pairing more difficult.  In the pairing approximation \cite{chica3}
part of the actual energy gap is due to the pseudogap and part to the 
superfluid gap. Thus the superfluid transition occurs only when the 
pseudogap alone is not enough to exclude the pairing. 
It occurs at a much lower temperature. Another effect may be related to 
quantitative differences between the quasi-particle and the self-consistent
calculations. The real part of the self-energy and the T-matrices are
 different in the two cases.  Below $T_c$ the self-consistent energy gap is 
always significantly smaller than the BCS result. This is different from  
the pseudogap pairing approximation \cite{chica3}, where at zero temperature
 the 
pseudogap goes to zero and the usual BCS result is recovered. We could not 
perform calculations at low temperatures, due to limitations in computational
resources. However, already in Fig. \ref{szcz}, a saturation of the 
energy gap in the self-consistent calculation is visible, around 
$\Delta(0)=7$ MeV.

For comparison we plot also the dispersion relation for the peaks in the BCS 
spectral function at $T=1.4$~MeV: $E_p=\pm\sqrt{\xi_p^2+\Delta^2(p)}$.
Since the energy gap is larger than in the self-consistent calculation the 
splitting of the two energies in the spectral function is much more pronounced.
It should be noted that the condition of weak BCS, $\Delta(0)\ll \xi_{p=0}$,
is not really fulfilled in the above case. It is another indication that 
weak coupling
BCS theory with free nucleon-nucleon interaction should not be used in the 
nuclear matter. 
The strong modification of the spectral function by the anomalous self-energy 
in the BCS case is visible also in the fermion momentum distribution 
(Fig. \ref{ferbcs}). The BCS fermion distribution is much broader than the 
Hartree-Fock distribution. Also the density, which is taken to be $\rho=.45
\rho_0$ for the BCS fermion distribution is $.43\rho_0$ for the Hartree-Fock
distribution.

\section{Conclusions}

We have presented a new way of performing nuclear matter calculations in the 
presence of nuclear superfluidity. The starting point is the T-matrix 
approximation for the nucleon self-energy. It allows to calculate the 
spectral properties of nucleons in an interacting system, treating on equal
 footing particle-particle and hole-hole diagrams. It leads also to
the formation of Cooper pairs below some critical temperature.
Strong scattering in the vicinity of the critical point, leads to a breakdown
of the quasi-particle picture, confirming the observation of \cite{ja}.
 This requires the use of self-consistent, and not quasi-particle,
fermion Green's functions for the calculation of the self-energy. 
A more important generalization of the formalism is required to
address the superfluid transition. In section \ref{pairing} such a scheme is
presented at least for the case, where the T-matrix approach indicates a 
second order transition.
 It consists in the use of off-shell propagators in the T-matrix 
ladder, one of them including also the  anomalous self-energy. 
In such a way the superfluid gap equation is equivalent to the condition
that the T-matrix has a singularity at Fermi energy and zero total momentum 
for all temperatures below $T_c$. The full nucleon spectral  function
includes the anomalous self-energy from the condensed pairs, and the 
normal self-energy due to nucleon-nucleon scattering. Both contributions are 
important for strongly attractive potentials in nuclear matter leading to an 
interplay of the pseudogap and superfluid gap around $T_c$.
The method presented in this work  has several important properties
\begin{itemize}
\item{} Above $T_c$, it leads to a self-consistent T-matrix resummation,
which allows for 
the description of the feedback of the pseudogap formation and 
of the scattering on the T-matrix.
\item{} It provides a link between
 the gap equation and the T-matrix equation, at the critical temperature 
and below. 
\item{} It gives the BCS theory in the limit of small scattering rates, 
unlike the approach \cite{hau1,peder}.
\item{} The value of the energy gap and the critical temperature are 
strongly reduced, in comparison to the BCS (quasi-particle) results.
\end{itemize}
When used with realistic nuclear forces, this approach presents
a procedure for calculating the nuclear matter properties, taking into account
self-consistently  single-particle spectral properties and  
fermion pairing. Both these important questions were not studied up to now
in actual nuclear matter calculations.
The first of these problems was discussed in a number of works related to 
high $T_c$ superconductivity \cite{chica1,pseudocm,sccm}; the first such
calculation being presented by Hausmann \cite{hau2}\footnote{Only Ref.
 \cite{sccm}
uses real time formalism for numerical calculations of the self-consistent 
T-matrix, like the present work.}.
The second question was addressed, to the author's knowledge, 
only by two groups
in condensed matter physics \cite{peder,chica3}. 
The work \cite{peder} uses a T-matrix resummation 
with both fermion propagators in the ladder including the anomalous 
self-energy.
However, due to the truncation of the resummation in the anomalous sector,
 the usual BCS gap equation is used. In that way, the gap equation is not 
consistent with the T-matrix resummation of the normal self-energy.
The approach \cite{d2} is similar, i.e. it uses two full propagators in the 
ladder diagrams. Thus, the gap equation is recovered only for small
gap, or is used in its linear form.
The work \cite{chica3} uses the pairing approximation, with one mean-field 
propagator and one with normal and anomalous self-energies. Both self-energies
 have  a BCS 
like form. 
Our approach is a generalization of this procedure,  using two
off-shell propagators in the T-matrix it
 allows us to use the off-shell propagator 
for
the calculation of the Hartree-Fock and scattering self-energies (\ref{hfsc}) 
and (\ref{imsc}), it is not restricted to a BCS like form of the normal 
self-energy, and it gives the most general T-matrix self-energy scheme above 
$T_c$. It should be stressed that there is no reason to neglect off-shellness
in one of the propagators in the T-matrix ladder. The consistency with the
 gap equation requires only to neglect the anomalous self-energy in one of
 these propagators.
The value of the pseudogap goes to zero at low temperature 
in the pairing approximation \cite{chica3}. On the other, hand the energy gap
in the self-consistent calculation is much smaller than the BCS gap also at
 low temperatures (Fig. \ref{szcz}). 

The numerical results are given for a very simple interaction. 
We plan to study the superfluid nuclear matter using more realistic separable
 interactions and also at zero temperature.
This would allow to  study  questions like the effect of 
pairing\footnote{The effect of the off-shellness of the propagators 
on the ground state energy has not been studied either.} 
interaction on the ground state energy, the interplay between pairing
in isospin $0$ and $1$ channels,
 and the pairing in neutron matter. Some more fundamental questions
remain also open. We use for the energy gap a mean-field approximation.
Moreover, the interaction potential in the gap equation is the free one.
Obviously modification of the pairing potential in medium, due to screening,
are expected. For consistency the same screened potential
should  be used in the 
T-matrix equation for  normal nucleon self-energy, meaning significant 
complications \cite{rep86}.
 Another, question is related to the cause of the reduction of 
$T_c$. In our calculation both the pseudogap and the nucleon scattering 
effects are included. To disentangle them the self-consistent calculation 
should be compared to results of the quasi-particle approximation and of  the 
pairing approximation below $T_c$.


\vspace{2cm}
This work was partly supported by the National Science Foundation
under Grant PHY-9605207.

\appendix

\section*{Appendix}

In this appendix we sketch the numerical method used in the
iterative solution of the coupled equations for the T-matrix, the
 the self-energy and the spectral function.
Special methods must be used for an efficient estimation
of  multidimensional
integrals, over energies and momenta, involved in the calculation of the
T-matrix and the self-energy when using off-shell propagators.

Let us enumerate below the steps of the iterative procedure, explaining 
in details the calculation of the energy integrals. We present the case 
of a separable potential of rank 1, as used in the actual calculations in 
this work. In that case the T-matrix in a given channel $c$ is given by
\begin{equation}
<k|T^{c \ +}(P,\omega)|k^{'}>= \frac{
\lambda_c g_c(k) g_{c}(k^{'})}
{1-\lambda_c J_c(P,\omega)} \ ,
\end{equation}
where the function $J_c$ depending only on the total momentum and energy 
is given by
\begin{eqnarray}
J_c(P,\omega) = 
 \int\frac{k^2dk}{(2 \pi)^2}\int d\cos(\Theta)\int \frac{d \omega^{'}}{2 \pi}
\int \frac{d \omega^{''}}{2 \pi} 
g^2(k) \nonumber \\
 \frac{A(p_1,\omega^{'}-\omega^{''})
A_s(p_2, \omega^{''})\big(1-f(\omega^{'}-\omega^{''})-
f(\omega^{''})\big)}{\omega-\omega^{''} + i \epsilon} \ , \\
p_{1,2}=P^2/4+k^2\pm P k \cos(\Theta) \ . \nonumber 
\end{eqnarray}
For rank one separable interaction, a nonzero solution of the  gap equation 
(\ref{intgap})
 is equivalent to the condition $1-\lambda_cJ_c(P=0,\omega=0)=0$
and the momentum dependence of the pairing gap is 
$\Delta(k)=\Delta(0) g_c(k)/g_c(0)$.
Let us present in the following, step by step,
 one iteration starting from the
calculation of the spectral function~:
\begin{enumerate}
\item{} Let us assume that the imaginary part of the self-energy is known
\footnote{In the first iteration a constant single-particle width  or
a stored array of ${\rm Im} \Sigma$ obtained earlier for a similar temperature 
and density is used.}.
The dispersive contribution to the real part of the self-energy can be
 obtained from  (\ref{disc}).
The calculation of the spectral functions $A$ and $A_s$ requires also the 
knowledge of the chemical potential $\mu$ and the pairing gap $\Delta$.
The condition for superfluidity is checked and if necessary the
gap equation (\ref{intgap}) is solved, using 
a numerical procedure for the solution of a nonlinear 
equation\footnote{This requires the calculation of the inverse T-matrix 
for a  particular value of momentum and energy 
(the gap equation) using a procedure to be described 
later.}. For each trial value of $\Delta$ the chemical potential and 
the corresponding Hartree-Fock energy are obtained with the constraint
(\ref{rhosf}) and using Eqs. (\ref{As}) and (\ref{hfsc}).
For each trial value of $\mu$ the coupled equations (\ref{hfsc}), (\ref{As}),
(\ref{spectral}), (\ref{rhosf}) are solved by iteration.
As a result we obtain the spectral functions $A$ and $A_s$.

\item{} In this step we calculate the T-matrix $T_c(P,\omega)$
for a range of values of  
total momentum and energy. 
Let us first calculate the imaginary part 
 of the function $J_c(\omega,P)$
\begin{eqnarray}
{\rm Im}J_c(P,\omega)=-\int\frac{d \omega^{'}}{2 \pi}\int \frac{k^2 dk}{8 
\pi^2}
\int d\cos(\Theta) A_s(p_2,\omega^{'})\big(1-f(\omega^{'})\big)
\nonumber \\
\label{nuj}
A(p_1,\omega-\omega^{'}) \big(1-f(\omega-\omega^{'})\big) g^2_c(k)
+ \dots \end{eqnarray}
the dots represent a similar term with factors $f$ instead of $(1-f)$.
The above $\omega^{'}$ integral is a convolution integral and can be performed
 using fast Fourier transform algorithms for numerical convolutions.
First the two factors in the energy integral are Fourier transformed
\begin{equation}
F_{(s)}(p,t)=FFT\left[A_{(s)}(p,\omega)\right] \ .
\end{equation}
Then the function 
\begin{equation}
{\bar J_c}(P,t)= - \int \frac{k^2 dk}{8 \pi^2} \int d\cos(\Theta)
F(p_1,t)F_s(p_2,t) g^2_c(k) \ ,
\end{equation}
is calculated using standard integration procedures for the two-dimensional
momentum integration.
The first term in Eq. (\ref{nuj}) is obtained by inverse Fourier transform of 
${\bar J}$. Analogously the second term in (\ref{nuj}) can be calculated.
The real part of $J$ is calculated using a dispersion relation
\begin{equation}
{\rm Re}J_c(P,\omega)={\cal P}\int \frac{d \omega^{'}}{\pi}
\frac{-{\rm Im}J_c(P,\omega)}{\omega-\omega^{'}} .
\end{equation}

\item{} The  imaginary part of the self-energy (\ref{imsc})
can also be written  as a sum of two convolution integrals
 in energy and can be calculated in 
a similar way as  ${\rm Im}J$ in the previous point.
This gives a next iteration of the single-particle width and the iteration
returns to the first point.

\end{enumerate}

A generalization to a separable potential of higher rank is obvious.
Since the Fourier transforms are performed before the loop over the total 
momentum, the most expensive numerical cost comes from the two dimensional
 integrals, when calculating ${\bar J}(P,t)$ for a range of values of $t$ 
and $P$. Very similar integrals must be performed when calculating the 
T-matrix and the self-energy in the quasi-particle approximation, 
except that here we have several such integrals.
In practice, when starting the iteration from a self-energy obtained in a
previous calculation at similar temperature and density, around 10 iteration
 are sufficient to converge with a relative deviation $\sim 10^{-4}$.

The numerical procedure in its present form introduces a cutoff in momenta 
of nucleons, to limit the energy range.
 Thus, we cannot treat more realistic potentials, whose cutoff
functions $g(k)$ involve large momenta.
 On the other hand, with decreasing 
temperature, the spectral function becomes narrow near the Fermi energy, and
cannot be  discretized on a finite grid of energy\footnote{The
 grid must be equally spaced 
in order to use fast Fourier transform algorithms}.
The extension of the present work to low energies and to large nucleon 
momenta requires a combined use of quasi-particle approximation near the
 Fermi energy and for large momenta and a continuum 
spectral function elsewhere, similar to the ansatz used in Ref. \cite{JL}.



\begin{thebibliography}{99}
\bibitem{nm} D.W. Sprung, in {\it Advances in Nuclear Physics}, edited by
M. Baranger and E. Vogt, (Plenum Press, New York, 1972);
C. Mahaux and R. Sartor, in {\it Advances in Nuclear Physics}, edited by
J.W. Negele and E. Vogt, (Plenum Press, New York, 1991)
\bibitem{fw} A.L. Fetter and J.D. Walecka, {\it Quantum Theory of Many-Particle
Systems} (McGraw-Hill, New York, 1971).
\bibitem{sfnm} L.N. Cooper, R.L. Mills and A.M. Sessler, Phys. Rev. {\bf 114}
 (1959) 1377.
\bibitem{d2} W.H. Dickhoff, Phys. Lett. {\bf B210} (1988) 15;
B.E. Vonderfecht, C.C. Gearhart, W.H. Dickhoff, A. Polls and 
A. Ramos, Phys. Lett. {\bf B253} (1991) 1.
\bibitem{vo} B.E. Vonderfecht, W.H. Dickhoff, A. Polls and A. Ramos,
Nucl. Phys. {\bf A555} (1993) 1. 
\bibitem{d1} B.E. Vonderfecht, W.H. Dickhoff, A. Polls and A. Ramos,
Phys. Rev. {\bf C44}  (1991) R1265.
\bibitem{ko} H.S. K\"{o}hler, Phys. Rev. {\bf C46} 1687 (1992) 1687.
\bibitem{roepke} T. Alm, G. R\"opke, A. Schnell, N.H. Kwong and
S. K\"ohler, Phys. Rev. {\bf C53}  (1996)  2181;
 A. Schnell, T. Alm and G. R\"opke, Phys. Lett. {\bf B387} 
 (1996) 443.
\bibitem{thouless} D.J. Thouless, Ann. Phys. (N.Y.) {\bf 10} (1960)  553.
\bibitem{km} L.P. Kadanoff and P.C. Martin, Phys. Rev. {\bf 124} (1961) 670.
\bibitem{ja} P. Bozek, nucl-th/9811073.
\bibitem{JL} F. de Jong and H. Lenske, Phys. Rev. {\bf C56} (1997) 154.
\bibitem{gw} M.I. Haftel and F. Tabakin, Nucl. Phys. {\bf A158}  (1970) 1;
M. Golberger and K. Watson, {\it Collision Theory} (John Wiley
 and Sons, New York, 1964).
\bibitem{kb} L.P. Kadanoff and G. Baym, {\it Quantum Statistical Mechanics} 
(Benjamin, New York, 1962).
\bibitem{paw1} P. Danielewicz, Ann. Phys. {\bf 152}  (1984) 239.
\bibitem{bm} W. Botermans and R. Malfliet, Phys. Rep. {\bf 198} (1990) 115. 
\bibitem{schmidt} M. Schmidt, G. R\"opke and H. Schulz , Ann. Phys. 
{\bf 202} (1990) 57.
\bibitem{yama} Y. Yamaguchi, Phys. Rev. {\bf 95} (1954) 1628.
\bibitem{baldo} M. Baldo, I. Bombaci, G. Giansiracusa, U. Lombardo, 
C. Mahaux and R. Sator, Nucl. Phys. {\bf A545} (1992) 741.
\bibitem{www} O. Benhar, A. Fabrocini and S. Fantoni, Nucl. Phys.
{\bf A550} (1992) 201.
\bibitem{pseudocm} M. Randeria, N. Trivedi, A. Moreo and T. Scalettar,
Phys. Rev. Lett. {\bf 69} (1992) 2001; P.G. McQueen,
D.W. Hess and J.W Serene, Phys. Rev. {\bf B 50}  (1994) 7304;
N. Trivedi and M. Randeira, Phys.
Rev. Lett. {\bf 75}  (1995) 312; 
R. Micnas, M.H. Pedersen, S. Schafroth, T. Schneider, J.J.
Rodr\'iguez-N\'i\~nez and H. Beck, Phys. Rev. {\bf B52}  (1995) 16223;
M.Y. Kagan, R. Fr\'esard, M. Capezzali and 
H. Beck, Phys. Rev. {\bf B57}  (1998) 5995; M. Randeira, cond-mat/9710223
\bibitem{peder} M.H. Pedersen, J.J. Rodr\'iguez-N\'i\~nez,
H. Beck, T. Schneider and S. Schafroth, Zeit. f\"ur Phys. {\bf B103}
(1997) 21.  
\bibitem{chica1} J. Maly, B. Jank\'o{} and K. Levin,
cond-mat/9805018; J. Maly, B. Jank\'o{} and K. Levin Phys. Rev. {\bf B56}
(1997) R11407; I. Kosztin, Q. Chen, B. Jank\'o{} and K. Levin,
Phys. Rev. {\bf B58} (1998) R5936.
\bibitem{chica3} Q. Chen, I. Kosztin, B. Jank\'o{} and K. Levin,
Phys. Rev. Lett. {\bf 81}  (1998) 4708. 
\bibitem{sccm} B. Kyung, E. G. Klepfish and P.E. Kornilovitch, Phys. Rev. 
Lett. {\bf 80} (1998) 3109.
\bibitem{hau2} R. Hausmann, Phys. Rev. {\bf B49}  (1994) 12975.  
\bibitem{hau1} R. Hausmann, Zeit. F\"ur Phys. {\bf B91} (1993) 291.
\bibitem{rsf} A.M. Sessler, in {\it Liquid Helium}, Proceedings of the 
International School of Physics Enrico Fermi, edited by G. Careri, 
(Academic Press, New York, 1963); V.G. Valeev, G.F. Zharkov and
Yu.A. Kukharenko, in {\it Nonequilibrium Superconductivity}, Proceedings
of the Lebedev Physics Institute vol. 174, edited by V.L. Ginzburg,
(Nova Science Publishers, New York, 1988);
D. Rainer and J.A. Sauls, in {\it Superconductivity}, Lecture Notes
of the ICTP Spring College in Condensed Matter on ``Superconductivity'', 
edited by P.N. Butcher and Yu Lu, (World Scientific, Singapore, 1995).
\bibitem{good} A.L. Goodman, Nucl. Phys. {\bf A186} (1972) 475.
\bibitem{strbcs} M. Baldo, U. Lombardo and P. Schuck, Phys. Rev. {\bf C52}
 (1995) 975.
\bibitem{migdal} A.B. Migdal, {\it Theory of Finite Fermi Systems}
(John Wiley and Sons, New York, 1967).
\bibitem{mahan}J.R. Schrieffer, {\it Theory of Superconductivity}
(W.A. Benjamin, Inc., Massachusetts, 1964);
G.D. Mahan, {\it Many-Particle Physics} 
(Plenum Press, New York, 1981).
\bibitem{anmo} P.W. Anderson and P. Morel, Phys. Rev. {\bf 123} (1960) 1911.
\bibitem{bawe} R. Balian and N.R. Werthamer, Phys. Rev. {\bf 131} (1963) 1553.
\bibitem{anisotropic} P.W. Anderson and W.F. Brinkman, in {\it
The Physics of Liquid and Solid Helium, part II}, edited by K.H. Benneman
and J.B. Ketterson, (John Wiley and Sons, New York, 1978).
\bibitem{rep86} A.D. Jackson, A. Lande and R.A. Smith,  Phys. Rep. {\bf 86} 
(1982) 55. 
\end{thebibliography}
\end{document}